\documentclass[epj,referee]{svjour}
\usepackage{graphics}
\begin{document}
\title{Observations of nonlinear eigen-states}
\subtitle{Including modal response of large spot size return from clattering contact structures}
\author{Michael C. Kobold\inst{1} 
\thanks{Dr. Pedro Encarnac\'ion of Colorado Springs and especially Michael McKinley of Arlington, Texas provided much appreciated assistance, www .linkedin .com /in/michaelmckinley2001 (remove spaces to activate link).}%
}                     
%
%
\institute{\emph{P O Box 32104 Panama City, FL 32407-8104} matadorsalsa at runbox dot com, 964 S. Shannon Ave. Indialantic, FL 32903. This is personal research. Otherwise, email the author at michael.c.kobold at navy dot mil.}
%
\date{Received: 0mar18 / Revised version: 0may18}
%
\abstract{
In scientific inquiry of small scale energy and transition events, the effect of an observer on the measurement can be substantial.  Similarly, the methods of observation can create entirely different understandings of the system behavior, discussed here on a macro scale.  An example using laser vibrometry analysis of nonlinear clattering contact of armor plate shows the difference between the common measurement-at-a-point compared to measuring the system dynamics from the simultaneous behavior of the entire structure. Using an optimum configuration for a laser vibrometer can improve remote identification of vehicles.  The former pencil-thin laser technique fits the keep-it-simple maxim but hides crucial coupling data.  With careful observation methods it is possible to measure the structural modal frequencies and calculate the transition probabilities between these discrete modes.  This work shows a genesis of what might be called 1/$f$ ``noise" which is explained herein by Zienkiewicz using modal analysis.  The  behavior also results from simplified analysis, and simulation -- that the energy of the essentially discrete structural vibration modes decreases with increasing mode number (and frequency).
} 
\maketitle
\section{Introduction}
\label{intro}

Using optics to sense vibration is a mature technology dating back to Foucault's methods of observations with knife-edge imaging to measure mirror shape in 1858.  Jean Bernard L\'eon Foucault invented a method to measure phase that was later used to remove unwanted phase effects.  ``\dots the Schlieren method, where all the spectra on one side of the central order are excluded." \cite{book:Born&Wolf} (This uses Foucault's knife-edge in the ``Fourier plane" at a distance of the focal length from a collecting lens.)  More recent physics discoveries use vibration to measure atomic size effects. Microscopy resolution smaller than the optical diffraction limit can be accomplished with the scanning tunnelling (Nobel prize) and other microscopes that allow images of large S-shells of atoms and even their P-shells.  The vibrating variant, the atomic force microscope (AFM) is able to detect separate molecules by measuring the difference in van der Waals force as a tiny cantilever scans over the sample, changing its resonance.  This article describes how the nature of observation can provide an entirely different picture of the vibrational modes of common vehicles and other integrated structures.

Three types of analysis provide support for this exposition of observation methods.  (i) Experimental results of a driven clamped bar with full fixity at both ends provides a laser vibrometry spectrum for both small and large probe beam spot sizes.  (ii) A master's thesis \cite{MSthesis:kobold06} by the author contains calculation results of imaging a vehicle hull at 4 kilometers. This allowed identification of the target vehicle based on the laser vibrometry spectral fingerprint.  The target armor clattering against the vehicle hull provides a surface vibration that modulates the probe beam. (iii) Several analytical calculations using explicitly nonlinear mechanics validate the clattering spectrum.  These experimental, simulation, and analytical results provide cross-supporting reinforcement for the three observational characteristics described in this paper.

This work defines three characteristics of composite structures such as vehicles.  1. The lower frequency modes are nearly discrete. 2) Energy from driven modes flows into other modes. And finally, 3) due to theory and observations defined herein, vibration strain-energy is higher in the fundamental mode, with energy decreasing as modal frequencies increase.  The latter two characteristics can be considered to be consequences of the second law of thermodynamics \cite{misc:meaningOfLife}.


An important issue to consider is how the probe beam spot size affects the observed vibration modes.  Some vibration modes may not be obvious from return due to a probe beam with a small spot size.  In the case of clattering armor plates, more mode information is acquired when viewing the entire optical image and processing its spectrum at multiple locations.  In spite of some interference, the non-imaging (spatially integrated radiant flux) spectrum of the probe return from the entire target's surface provides more information than the small spot size in that phase information over time is embedded in the sensed signal.


\subsection{Three observational characteristics}

\begin{figure}
\resizebox{9cm}{!}{
\includegraphics{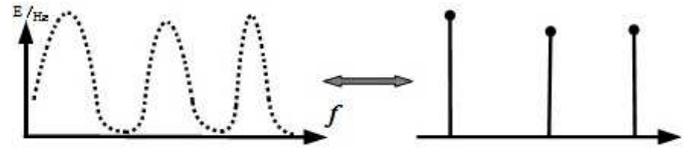}}
\caption[discreteModes]{\label{discreteModeSketch}At the fundamental frequency and a few higher modes the resonance quality \textit{Q} is high -- for systems without large damping the ``lower" modes (see Figure \ref{modalOrderSketch}) are effectively discrete modes.}
\end{figure}

\begin{figure}
\resizebox{8.5cm}{!}{
\includegraphics{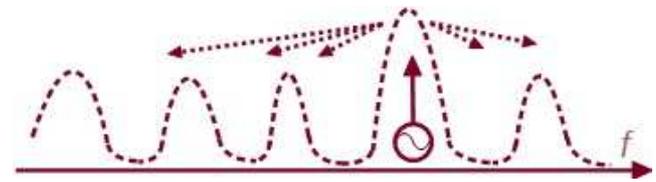}}
\caption[transferEnergy]{\label{energyXfrSketch}Vibrational strain-energy from different components transfers from driven modes (power symbol, slanted $\ominus$) to lower energy modes at different frequencies.  The nonlinearity of all common, practical fasteners (bolted and riveted joints) allows energy transfer.}
\end{figure}

\begin{figure}
\resizebox{8.7cm}{!}{
\includegraphics{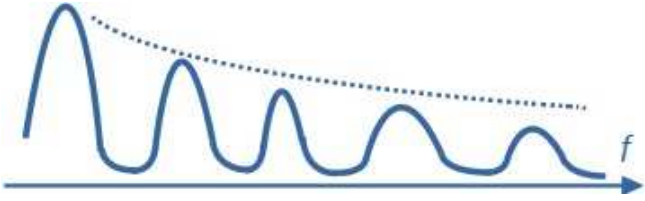}}
\caption[modalInverse]{\label{modalOrderSketch}As described by the Zienkiewicz quote in Section \ref{ZienkiewiczEnergyFlow}, high frequency modes eschew energy in favor of lower frequency modes.  Therefore the lower modes ``fill up" first but also retain more energy than higher modes unless the lack of vibration coherence can block the transfer or interfaces that act like active systems manipulate the energy.}
\end{figure}

Based on the three previously mentioned supporting results, the experiments, simulation, and analytical results, three observations become apparent:  (1) the lower frequency modes have sharp resonances that are effectively discrete responses Figure \ref{discreteModeSketch} that can be modeled on center frequencies when damping is limited to structural damping \cite{inProc:gByTedRose01}, as we seen in Figures \ref{missedMode} and \ref{allModes}.  (2) energy transfers from one component and its modes into the lower modes of nearby components through joints, including welds, which have nonlinear load-deflection curves.  Each mode is comprised of a 3-D mode shape (deflection vectors) of the entire structure (Figure \ref{energyXfrSketch}).  The strain-energy (modal energy) of each mode couples to lower energy modes when and if the spectrum reaches steady state.  Finally, (3) modal energy flows into the lowest frequencies first (Figure \ref{modalOrderSketch}), depleting the energy flow as the mode number increases from $f_0, f_1, f_2, \cdots$ and up through the closely-spaced high frequency modes that are practically infinitely dense, although for discrete modes \emph{FEA} is limited to the number of degrees of freedom (DOF) in the model.

The energy at each of the modes shown in Figure \ref{discreteModeSketch} transfers to lower energy modes if physical load paths exist that are conducive to mode coupling.  A transfer of energy from a strongly driven mode to other modes appears in Figure \ref{energyXfrSketch}.  That driver might be in a component located far from the receiving modes, but in `frequency space' they might be coupled.  It might be driven by the engine  at one frequency, yet sensed in the back seat at another frequency.  Nonlinearities in response create harmonics of resonances that allow energy to couple with lower modes.  The the low energy tails of wide resonances (damped broadening) also touch other lower modes.  Another energy transfer avenue is clattering where spatial distribution of mode amplitudes causes contact that excites different modes, much like plucking a musical string at different locations can change the tone.  The concept of energy transfer between states relies on the frequency of the mode, which is the square root of the eigenvalue in an analysis of normal modes.

Figure \ref{modalOrderSketch} is a sketch where the final steady state of the vibrational system has energy ordered from strong resonances at the fundamental frequency to a paucity of energy at higher frequencies. A quote from Zienkiewicz in Section \ref{ZienkiewiczEnergyFlow} provides an explanation for 1/\textit{f} dependence of modal energy.

The models, structures, and observational characteristics that comprise this article are summarized in the Conclusion in Table \ref{TableI}.

\subsection{Physical model}

\begin{figure}
\resizebox{8.4cm}{!}{
\includegraphics{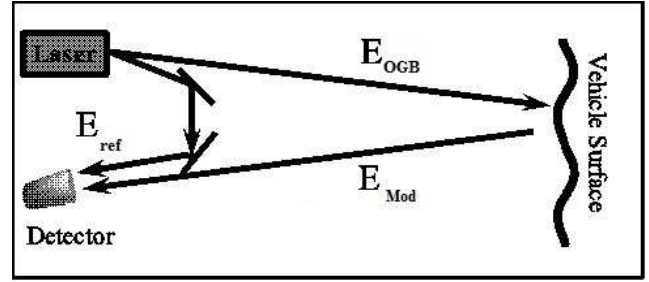}}
\caption[Laser Vibrometry]{\label{basicSkematic}Coherent irradiance undergoes a spatially harmonic phase modulation as a function of the probe beam location due to target vehicle surface skin structural vibration.  $E_{OGB}$ is the outbound Gaussian beam, $E_{mr}$ is the modulated return, and $E_{ref}$ is the reference beam for this simulation model}
\end{figure}

While conventional laser vibrometry systems can better identify many modes in an academic structure - such as this solitary vibrating bar in the following photos - by using a ``pencil-thin" beam, a large spot size can adequately measure those modes and more.  This can be shown by simulation and analytical calculation. The simulation uses finite element analysis (FEA) for vibration deformation data that is handed-off to a set of MATLAB\textsuperscript{\textsc{tm}} functions that perform Fresnel propagation for the optical sensing. The FEA shows the mode shapes and calculates the energy per mode which are the modal participation factors (MPFs).  The FEA surface vibration data is input for the MATLAB\textsuperscript{\textsc{tm}} image propagation calculation (Fresnel propagation \cite{book:GoodmanFO68}).  The physical schematic for the simulation appears in Figure \ref{basicSkematic}.  Laser vibrometers use many different methods that can have an affect on the precision of the measurements.\footnote{This interferometric model allows analysis of the phase modulation of the exitance from the vehicle (the \emph{return}), hence its use for the optical analysis.  At some level, even the portion of laser vibrometers that use Doppler to measure vibration have images that are affected by phase modulation generated by vibration mode shapes.  This is, in fact, part of the mechanism of spectral elimination.  Phase modulation is used here to determine the effect of mode shapes on the return for both imaging and non-imaging systems.}



The laboratory bar measurements show the signal to noise ratio (SNR) is quite high for structural components and, unless the system is purposely damped, the resonances are sharp.  These sharp resonances lead to the discrete mode picture that the field of modal engineering is built to exploit.  The literature provides the theoretical reasons that strain-energy tends to eschew higher frequency modes and thus populate the lower modes.  Years of experience of vehicle design and analysis support these conclusions, as does the FEA of clattering armor plate.  The FEA results show that unsymmetrical modes of the clattering plates have higher frequencies than symmetrical (in-phase) vibration of the parallel plates, and that these modes are discrete below 500 Hz, as is the case with typical automobiles.  These modes are usually easy for laser vibrometers to detect.  The optical image propagation model shows that even a non-imaging spectrum of the surface of the clattering plates contains energy-ordered modes whose vibration strain-energy drops off with frequency, as described in the Zienkiewicz quotation of Section \ref{ZienkiewiczEnergyFlow}.

Commercial laser vibrometers use pencil-thin spot sizes on the object being measured.  This is equivalent to one pixel's response from a large spot size imaging system.  Non-imaging signals average responses from all pixels for each time step.  An optimum automated target recognition (ATR) system could use a combination of small and full illumination.

\section{Structural nonlinearities}

\label{structNonlinSection}

A vibrating surface can be measured for dynamical properties by using any of several forms of laser vibrometry.  The type of laser vibrometer is less important than the phenomenology of what can be measured remotely.  A priori project restraints such as requirements to calculate a transfer function, or to use pencil-thin probe beams, can limit the sensed spectra and thus omit or change the perceived vehicle behavior from reality in a manner unrelated to measurement or analysis error.  Especially for complicated nonlinear systems, it sometimes pays to ignore detailed kinematic analysis, which are often invalid for nonlinear systems, in favor of \emph{modal engineering} which tabulates the modes (deformed shapes for particular resonance frequencies), their frequencies, and their energy or \emph{modal participation factors}.



For over half a century modal engineers have routinely done modal analysis for nonlinear vehicle structures.  This work uses a structural model of a plate bolted to another plate on a base structure.  The work also uses simplified and even ``linear" component models, but they cannot deliver the behavioural metrics that the full model provides.  They merely verify limited aspects of the nonlinear model, which is a crucial part of the overall analysis.  In the main analysis, the structure is free-floating system using D'Alembert reactions, an analysis technique common to vehicle engineering.  See, for example, the numerous `quarter-panel models' in the literature \cite{ppr:mitra06,misc:koboldExperience}.

The sensor model is an optical model without the usual air turbulence analysis, except to check for Fante's wavefront coherence breakup range \cite{ppr:breakupFante75} and a few other imaging issues \cite{MSthesis:kobold06} that are out of the scope of this paper.

The physical model for the clattering armor assumes a linear superposition of vibration modes on a nonlinear structure.  Some effects that are ordinarily assumed linear for structural dynamic response must be nonlinear at the outset.  For one, this is not a linear time-invariant system.  See the details of removal of these typical LTI assumptions and its relationship to stability, stabilization, and stabilizability in the thesis \cite{MSthesis:kobold06}, available through DTIC dot mil, also contain a treatment of how fixity changes the modes, which can model aging of parts.  Therefore, the modulation of the sensed optical radiant flux is related to the analytical nonlinear expressions in the Appendix that help explain the contact nonlinearity in the FEA simulation.  The modeling complications related to fixity, nonlinearities, and modal participation factors are dependent on control of the nonlinear solution per time step and over time.  The analyst also needs to make sure the solution follows the load deflection curve \cite{book:Zienkiewicz91}.

The finite element analysis (FEA) output represents the three-dimensional structural model of the clattering armor plate system and provides the time history of the structural vibration that modulates the optical return recorded by the sensor.  The Appendix contains a set of single and two degree of freedom analyses result in analytical expressions that provide insight into the FEA results.  Discussions concerning the need to have load increments, iterations, and other aspects of the FEA solution follow the millions of degrees of Freedom (DOFs), each with their own load deflection curve, are found in the author's thesis \cite{MSthesis:kobold06}.  The objective of the analyses is to use the optical images to determine the frequency of structural spectral modes for target identification (ID).

In recent conversations with Navy scientists and engineers the author found himself describing how the nature of system observation plays an important role in perception of the modal system and how the low frequency strain-energy modes of the structure are quantized.  The discrete nature of the modes has been an underlying assumption of modal analysis and in its application to ride-and-handling for over 20 years.  The modal participation factors describe the dynamics of energy flow by using vectors of MPF's that span the pertinent modes.  Each MPF is an element in the $ \mathbf{\Phi[\mathit{f_i}]} $ vectors, arrays ordered by mode number \textit{i}, used in the FEA of vibrating structures such as vehicles, antennas, and spacecraft.  For example, NASTRAN\textsuperscript{\textsc{tm}} may require `DMAP alters' (macros) to read out some of the components of $ \mathbf{\Phi[\mathit{f_i}]} $.

An example of the unintended consequences of design decisions for the 1990 era Corvette follows this detailed restatement of the three essential system characteristics introduced earlier.  They will be used in the example.  (1) The lower frequency modal states (eigenvectors) form discrete modes even for complicated system synthesis models and the vehicles they represent.  The work of vehicle design for these issues focuses on (2) transition probabilities\footnote{Transition probabilities are related to structural coherence spectra by modal engineers who seek to reduce undesired vibration modes (resonances) in the structure.\cite{ppr:AllemangMAC03,phdth:Allemang80}} and (3) the flow of strain-energy that preferentially fills the lower modes with energy, as discussed in the quoted passage in the next section.

\textbf{Three observational characteristics of modal systems:}

1. The strain-energy levels within each modal state are tabulated by the structural design team in order to determine critical components.  These critical regions absorb most of the ride-and-handling and structural engineering effort during vehicle design, resulting in a final list of lower frequency energy levels\footnote{Acceleration spectra are measured in units such as strain (related to strain-energy amplitudes) or acceleration per hertz.  But in practice the ``power" is displayed, signal processing ``power'' being the square, element-by-element, in $g'^2$/Hz.  However, the optically sensed units are dB/Hz.}  The lower modes relate to modes shapes that are usually characteristic of the structure, such as bending, twisting, and stretching.  A real-life example appears later in this section.

2. The transitions between modal states (eigenvectors of the structure) are often sigmoid in nature\footnote{While there are systems where joints change behaviour on a short-term basis during operation (e.g., magneto-rheological and similar semi-active systems), the sigmoid transition is usually driven by two factors: (A) energy thresholds (e.g., bolted joints have a bi-linear load curve \cite{man:aeroStrNASA}), and (B) design changes during the drawing release phase of vehicle development.} where the region of rapid mode coupling depends on meeting strain-energy thresholds that allow more efficient coupling between modes.  Most of the energy tends to flow to components with lower fundamental modes according to characteristic (3) below.  Nearly all joints are structurally nonlinear\footnote{Even spot welds have at least geometric nonlinearities in the Green's strain tensor. This is due to large strain related to dissimilar stiffness and offset load moments (`kick's' in aerospace jargon).  Nonlinearities are usually due to frictional hold of bolted or riveted joints that ``\emph{must be ignored}" in most structural analyses \cite{book:Popov98}.  Therefore, most system FEA cannot model MPFs unless the model is nonlinear and designed to develop MPFs.} for the transfer of forces and reactions, a situation that provides substantial coupling nonlinearity.

3. The low frequency end of the spectrum tends to hoard the majority of the strain-energy, as the quote from Zienkiewicz explains.

These characteristics are summarized in the table \ref{TableI}, the Conclusion.  Analysis of MPF's (developed by characteristic 2) can determine state transition probabilities.  However, modal engineers tend to be focused on solving critical failure issues.  Engineers of any kind might use a different kind of `mode' for `failure mode effects analysis' (FMEA).  However modal engineering involves creation of a record of the order of the vibration modes, followed by an analysis of the tracks of the modes as the design variable change.


A structural vibration example described below shows how a structural design change can de-couple modes in order to improve ride-and-handling.  However, this change has consequences other than aesthetic design constraints.  The changes can increase crashworthiness risk and other seemingly unrelated systems, and they can degrade previously adequate vibration and modal engineering balances in the design.  For example, stiffening one joint or component allows for a much more efficient flow of energy through that element of the structure.  Strain-energy is drawn from the high frequency modes to the low modes in a surprisingly efficient manner once such a structural ``channel" is open (components have structural coherence).  Then the MPFs start to re-balancing as the entire integrated structure starts to equilibrate.  So a stiffener can solve one problem and cause another.  The engineering group that deals with braking and crash loads that transfer from the front to rear bumper can have its margins go from adequate to negative due to a design change by the ride-and-handling team that solves a mode-coupling problem.  This is similar to an old solution to convertible mode coupling:

\subsection{Inter-organizational effects of design changes}

The 1990 era Chevrolet Corvette convertibles an initial design had a fundamental bending mode just below 30 Hz, which put it very close to the suspension mode.  This is a overall vehicle bending mode that curves about a lateral axis, where the bumpers move vertically and together, both in direct opposition to the vertical motion of the seats. It is usually the lowest frequency vehicle mode for a convertible as (if the vehicle is trying to do situps).  The Corvette designers needed to stiffen this mode to reduce coupling to the nearby suspension modes; they were also near 30 Hz. The models in the late 1980's had very tall rockers that satisfied this requirement.  This was the most efficient structural fix because the stiffness of a beam to bending vertically is proportional to the cube of its section height \cite{book:Roark89}.  Apart from the aesthetic issue with having to step high to get into the Corvette, these tall rockers also changed other load paths including a for-aft transmission of load, important to crashworthiness, as well as making the vehicle slightly more heavy.  Automotive is one of the few industries that has a price on the engineering sufficient to remove a kilogram of mass, which was \$50,000 in 1990.  These and other costs were accepted in order to move the fundamental bending mode away from the suspension mode to effectively eliminate coupling.  For later years, other technologies helped solve this issue.

With the exception of the Pininfarina chassis test results \cite{inProc:Pininfarina}, few modal analyses such as these (chassis or body-in-white) are in the literature.  To a scientist, the ride-and-handling issues and how they relate to the three observational characteristics described above might seem to be basic enough to warrant several scholarly articles.\footnote{Physicists in Ann Arbor were baffled surprised that a wheelbase change of less than 2 cm would require over a year of mechanisms analysis to re-balance just the toe, caster, and camber - in negotiations with competing requirements from groups that engineer tires, turn radius, vibration, and crashworthiness.}  However, this is clearly the arena of trade secrecy at such a high level of value that corporate lawyers would be unsurprised at the paucity of measured data available to the public.

\subsection{Measured data: Vibrating clamped-clamped bar}

\begin{figure}
\resizebox{8.4cm}{!}{
\includegraphics{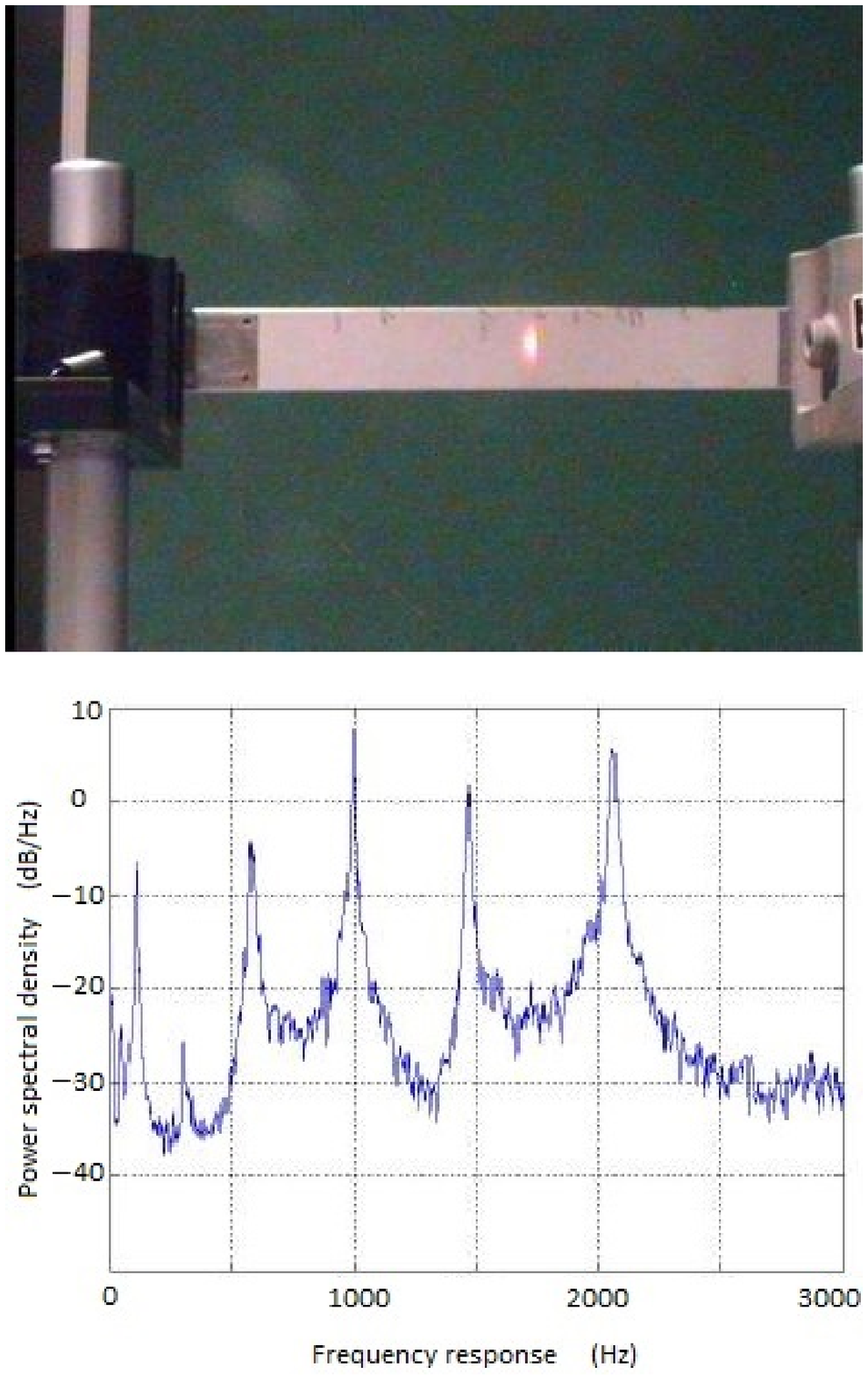}}
\caption[Spectral Elimination]{\label{allModes} The image from a small spot size makes a spectrum with a $f_3$ = 1460 Hz mode. Permission N. Pepela \cite{MSthesis:NgoyaPepelaLaserVib03,ppr:insensitive}.}
\end{figure}

\begin{figure}
\resizebox{8.4cm}{!}{
\includegraphics{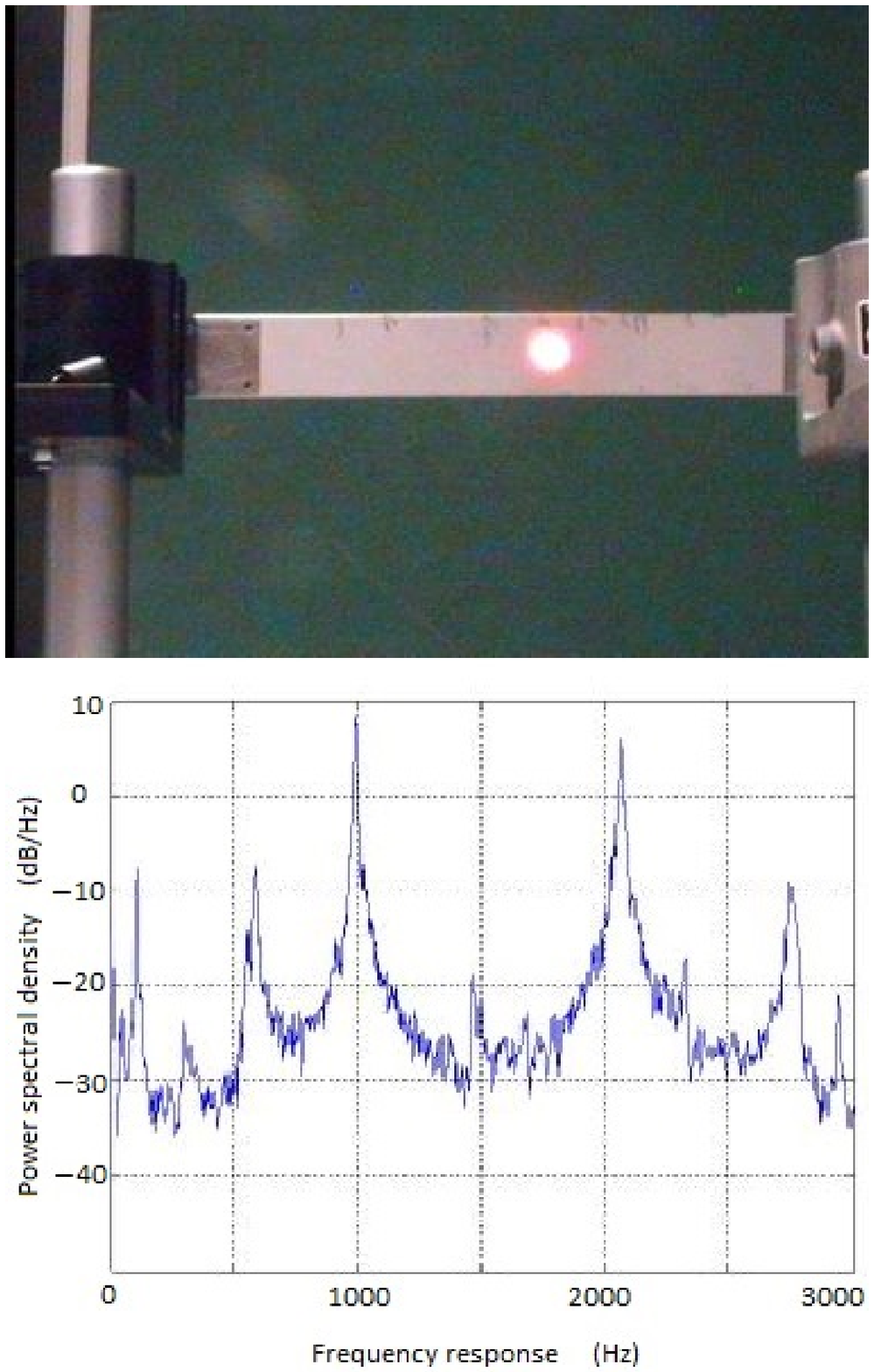}}
\caption[Spectral Elimination2]{\label{missedMode}A larger laser spot size creates a laser vibrometry spectrum where the $f_3$ = 1460 Hz mode phase-averages away.}
\end{figure}

Pepela showed in his thesis \cite{MSthesis:NgoyaPepelaLaserVib03} that the modes are  not just maxima of a spectrum, but spikes with huge SNRs in the spectral response.  Figures \ref{allModes} and \ref{missedMode} from his thesis \cite{MSthesis:NgoyaPepelaLaserVib03} each show a photo of the vibrating bar illuminated by a laser vibrometer with the corresponding vibrometer spectral density plot shown in the lower portion of each figure.  The first (Figure \ref{allModes}) shows all the major modes whereas the second (Figure \ref{missedMode}) is missing the modes at 1460 Hz.  In Figure \ref{missedMode} the vibrating bar is seen to be illuminated by a laser beam with a large round spot size centered on the 1460 Hz node of the vibrating bar.  In Figure \ref{allModes} the right half of the laser beam of Figure \ref{missedMode} has been blocked, illuminating only the left half of the 1460 Hz node.  The halved beam size (Figure \ref{allModes}) allows the laser vibrometry spectral analysis system to show a stronger 1460 Hz mode due to the elimination of spatial averaging introduce by phase-related destructive interference from both sides of the 1460 Hz node.


Initially spectral elimination might appear to be a problem for the laser vibrometry industry for use with non-imaging sensors if the spot size is large.  However, Pepela's thesis \cite{MSthesis:NgoyaPepelaLaserVib03} and another article by the author \cite{ppr:insensitive}, with support from simulation and analytical calculations, show numerically how unlikely spectral elimination is for commonly manufactured items, even if the spot size encompasses the entire vehicle.  Furthermore it is necessary, but not sufficient that super-symmetric structures produce spectral elimination in laser vibrometry \cite{MSthesis:kobold06}.  The vibrating bar is the simplest form of structural super-symmetry that results in spectral elimination, but it is rarely the only structural component a laser vibrometer can use for identification of a manufactured structural system (vehicle), unless the target is the vibrating bar (a structural \emph{oscillator}) that requires identification.  But even for such oscillators, the typical transducer is a piezoelectric system that is not super-symmetric, and thus not prone to spectral elimination; it is not a driven bar.  Therefore, a vehicle that is purposefully fitted with vibrating bars can still be identified with its spectral ``fingerprint" in spite of possible spectral elimination from oscillators.  Pepela's result \cite{MSthesis:NgoyaPepelaLaserVib03} was that use of small spot sizes reduces spectral elimination from the laser vibrometry spectral result (Figure \ref{missedMode} vice \ref{allModes}).

Michael McKinley of Arlington, Texas, observed that Figure \ref{allModes} contains modes (at approximately 2740 and 2900 hertz) that are not visible in the small spot size collection in Figure \ref{missedMode}.  While the small spot size is still large enough to average out higher frequency (smaller vibration shape wavelength) modes, another spectral elimination that small spot size vibrometers are susceptible to is the effect of the probe spot being on a Chladni line of nodes \cite{book:RayleighTOS1} of the vibration shape for particular frequencies \cite{book:ChladniWiki}.  Detection of these modes at 2300, 2740, and 2900 Hz is a positive observational difference provided by using a large spot size, in this case.

\subsection{Observational differences}



The difference between small spot size (Figure \ref{allModes}, and larger spot size (Figure \ref{missedMode}) comprises one dimension of observation variation that can change the observer's view of reality.  Physical systems such as the Corvette design example can be used to explain the existence of the three characteristics of modal systems listed above.  Contact nonlinearity is the main physical model of interest in this paper because contact is a relatively simple and ubiquitous form of nonlinearity.

\section{Zienkiewicz' energy ordering proposition}

\label{ZienkiewiczEnergyFlow}

Vibration strain-energy transfers from one component to another, usually through joints that are necessarily nonlinear (even spot welds), depending on the amount of `fixity' of that interface or joint.  An extensive treatment of how fixity changes the modes and therefore the sensed optical radiant flux is in the author's thesis \cite[B.2.3, p. 187]{MSthesis:kobold06}, including nonlinear examples that compare to the contact nonlinearity of the FE model.  Fixity is an engineering variable for a particular DOF that determines the ratio from zero to one (0\% to 100\%) of force, moment, or torsion \cite{book:Roark89} that will transfer from one component to another through the joint member for whom the fixity is defined \cite{book:Popov98}.

Even ``simple" vehicle structures have complicated nonlinear load paths (mostly through joints) where forces from one region of the vehicle affect parts nearly everywhere else on the vehicle.  These are the pathways for vibration energy flow that tend to dump energy into low frequency structural modes.  Modal engineers use isolators and suppressors to stymie this natural tendency, but mostly just for critical components. Some components tend to have non-negligible vibration spectral energy down at these ``fundamental" and near fundamental modes\footnote{The `fundamental' mode is the one at the lowest frequency.} even without the nonlinear effects. Damping helps drive the energy into the lowest modes because the resonance spreads (widens) to encompass a larger bandwidth.

Zienkiewicz showed  \cite[340-341]{book:Zienkiewicz91} energy transfer using only viscous damping to form a ratio of damping to its critical value,\footnote{Critically damped systems remove all oscillation.} $c_i = 2 \omega_i c_i' $ as quoted below.  For vibration systems that are driven by a forcing function \textit{f} Zienkiewicz describes the DE  $ \mathbf{M\ddot{a} + C\dot{a} + Ka + f = 0} $ using a mass matrix ($\mathbf{M}$), a damping matrix ($\mathbf{C}$), and a stiffness matrix ($\mathbf{K}$).

\textbf{Zienkiewicz quote:}

\begin{quotation}
`` ... we have indicated that the damping matrix is often
assumed as $ \mathbf{C} = \alpha \mathbf{M} + \beta \mathbf{K} $.
Indeed a form of this type is necessary for the use of modal
decomposition, although other generalizations are possible
[references given in the book].  From the definition of $ c_i' $,
the critical damping ratio [described above], we see that this can
now be written as

$ c_i' = \frac{1}{2\omega_i}\mathbf{\overline{a}}_i^T
(\alpha \mathbf{M} + \beta \mathbf{K}) \mathbf{\overline{a}}_i =
\frac{1}{2\omega_i}(\alpha + \beta\omega_i^2)
$

Thus if the coefficient $ \beta $ is of larger importance, as is
the case with most structural damping, $ c_i' $ grows with  $
\omega_i $ and at high frequency an over -- damped condition will
arise.  This is indeed fortunate as, in general, an infinite
number of high frequencies exist which are not modelled by any
finite element discretizations." \cite[340, 341]{book:Zienkiewicz91}
\end{quotation}

Therefore, vibrational strain-energy tends to naturally migrate from higher to lower frequency modes.  For linear systems some of the strain-energy can find its way down to the lower modes because over-damping causes more overlap of modal response resonances.  Nonlinearities add to the methods of energy transport as discussed later.

This results in a set of modes where the mode number increases monotonically with frequency and has an energy that decreases monotonically with mode number.  Possible exceptions to this natural state include artificial structures\footnote{Examples of artificial exceptions to the 1/\textit{f} phenomenon could be exotic.  Perhaps a feedback system might ``manually" excite high harmonics without exciting lower harmonics, such tapping very near the base on a taut chord would tend to excite the higher frequency mode.} or temporary transients such as might be introduced with magneto-rheological fluid or other systems controlled by magnetic or electric field variations.  The 1/\textit{f} phenomenon is found in many different fields of engineering.

The local mode structure in Pepela's laboratory test measurement in Figures \ref{missedMode} and \ref{allModes}, which do not follow the 1/\textit{f} phenomenon within the displayed 3 kHz,  \emph{appears} to be a contrary example. This an example of not being able to `see the trees' within its part of the response spectrum of the very stiff clamped-clamped structure.  For those results, the vibrating bar only had a half-dozen distinct modes across 0-3000 Hz, one of which was diminished in Figure \ref{missedMode}.  The utility of the structurally super-symmetric bar is that it shows discrete modes, and that the energy in some modes is reduced by spatial ``averaging" -- integration over the area of a surface of variable phase that introduces varying levels of continuous interference depending in large part on the limits of spatial integration \cite{ppr:insensitive}.  But is this really a counter-example of the 1/\textit{f} phenomenon?

That it is not becomes apparent by analysis of the physics of a simple model of a spring of stiffness \textit{k}.  This explanation considers multi-variate non-uniform parametric change in the system.  The maximum potential energy is approximately $U_{\mbox{max}} \approx kx^2/2$ where \textit{x} is the largest extension of a simple spring.  Assume that the frequency $ f^2 = k/(4 \pi^2 m)$ is increased, then the energy grows as the square of the frequency, parabolically as $U \approx 2 \pi^2 mf^2x^2$.  In practice, the deflection \textit{x} decreases with frequency, as it must.  However, there is a practical limit to reduction in deflection.  When that limit is reached, while the frequency continues to increase, the potential energy must grow.  This means that higher frequency modes require more energy, even for smaller deflection -- especially when deflection is already small, when $x^2 << U / (2 \pi^2 mf^2)$.  The kinetic energy side of this simple calculation is \emph{Rayleigh's method} \cite{incoll:MarksHdbk79}.

It may help to see this from a force point of view using the gradient of $U$.  If the restoring force is conservative, it is $F = \partial U / \partial x = 4 \pi^2 mf^2x$, and is thus a quadratic function of the frequency, $f$.  It might be tempting to consider a ``conjugate force" for frequency, where the gradient of $U$ is $4 \pi^2 mx^2f$ (units of action : Energy $\times$ time, or momentum $\times$ distance) \cite{book:CRChdbkChPh}.  However, the force ($F = \partial U / \partial x$), cannot increase ad infinitum.  At some point assumptions of linearity and structural integrity start breaking down.  Let us combine the observations of this upper limit on the force,and the understanding that the spring extension is limited to $x > 0$, the proposition that the energy required to have a particular mode increases approximately quadratically, ``on average" in this sense of the other variable being limited, and experience that dynamical systems naturally tend to avoid populating the higher modes with energy.  Many children experience this when they try to forcefully excite a higher frequency mode in a rope, only to fail unless stressed to the point of overexertion.

\begin{figure}
\resizebox{8.8cm}{!}{
\includegraphics{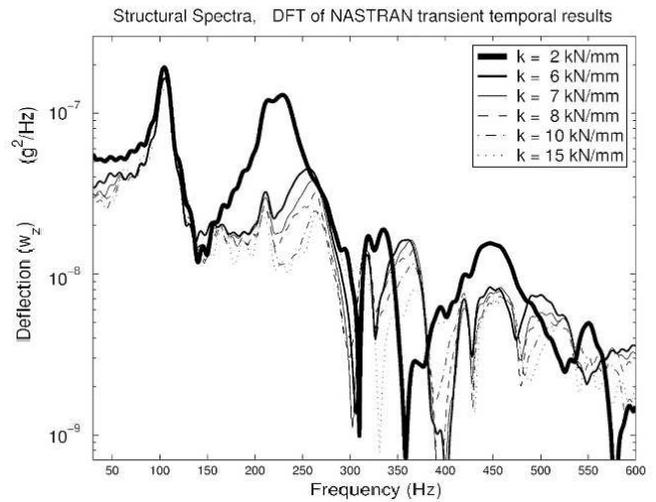}}
\caption[structuralSpectrum]{\label{impactResponseFRF} The spectrum of several armor-hull clattering systems except for the symmetrical modes.  The results for several with different armor-hull baffle stiffnesses (see legend).}
\end{figure}

A conjecture can be stated using the theory above for the system restoring force of simple harmonic motion, based on the gradient of potential energy, along with the FEA simulation included herein (Figure \ref{impactResponseFRF}), including industrial experience, and the test data from the childhood of most of us, as described above.  it is reasonable to assume the conjecture that, while system energy decays with frequency ``on average," as described above, a plot of the spectrum within its lowest half dozen modes might not show the 1/$f$ phenomenon, but that the full pattern dues.  Sometimes the spectrum plot is `in the trees' and the ``forest" of the 1/$f$ phenomenon over the entire large bandwidth spectrum is not seen for those ``trees."  Consider the `in the trees' ranges of 150-250 Hz and 400-500 Hz for the much lower fundamental frequency system of clattering armor in the FEA results of Figure \ref{impactResponseFRF}.

This plots several spectra of the artificial structure, clattering armor.  There is one curve for each level of damping (see legend) and they show a \emph{different} method of inter-modal energy transport.  The 1/$f$ pattern shows up in its structural vibration spectrum.  A collection of the deformed shapes for transient analysis in the CSC thesis \cite{MSthesis:kobold06} shows how the transient result at each time step is a superposition of normal modes (eigenvectors) populated with energy according to the MPFs $\Phi(f)$. Looking at the normal modes displayed in the CSC thesis it becomes apparent how one mode at one frequency for one plate excites another mode at a different frequency on the other plate because of deflections that line up along the surface to contact the other plate at anti-nodes of a different mode. The lower modes tend to be receiving modes.


Using appropriate structural damping, the frequency response function in Figure \ref{impactResponseFRF} is the result of an impulse load that rings all the modes of the FE model.  It appears to be natural that MPFs decrease with the mode number of the mode they measure; the vector is an array of monotonically decreasing participation values (in a time-averaged sense), usually measured as the strain-energy for each mode.  This is part of the reason for the ubiquitous 1/$f$ phenomenon.  At least for structural vibration, \cite{book:FEAbyMacNeal94}, we can thank the Zienkiewicz section quoted above \cite[340, 341]{book:Zienkiewicz91} for a partial explanation of this 1/$f$ fall-off in the quote above.

\section{Simple harmonic motion, not so simple}

\label{SHO}

A spring-damper-mass \emph{single} DOF system (SDOF) provides a one dimensional model of the clattering armor plate which can provide analytical solutions to the vibration DEs.  Assume `small deflection,' where strain-energy density is low enough for linear elasticity assumptions (lack of permanent set in the structure being modelled).  Also assume that modes are nearly monochromatic functions of sines and cosines.

In a more formal development of the transfer of energy from one mode to another, Lord Rayleigh points out that pure sine or cosine waves do not exist in reality.  He derived the differential equations for Newton's theory for isothermal compressive-vacuum vibration resulting in this second order form \cite{book:RayleighTOS2}.

$ \bigg(\frac{dy}{dx}\bigg)^2\frac{d^2y}{dt^2} =
\frac{dp}{d\rho}\frac{d^2y}{dx^2} $

Lord Rayleigh then comments on the ability of simple harmonic motion to maintain shape in nature.\footnote{``Since the relation between the pressure [$p$] and the density [$\rho$] of actual gases is not that expressed in [ $ p=const-(u_o^2\rho_o^2 / \rho) $], we conclude that a self-maintaining stationary aerial wave is an impossibility, whatever may be the velocity $ u_o $ of the general current, or in other words that a wave cannot be propagated relatively to the undisturbed parts of the gas without undergoing an alteration of type.  Nevertheless when the changes of density concerned are small, [ $ p=const-(u_o^2\rho_o^2 / \rho) $] may be satisfied approximately; and we can see from [ $dp/d\rho=(u_o^2\rho_o^2 / \rho^2) $ ] that the velocity of stream necessary to keep the wave stationary is given by [ $ u_o=\sqrt{dp/d\rho} $ ] which is the same as the velocity of the wave estimated relatively to the fluid." \cite{book:RayleighTOS2}}  The extent to which the mode shapes are not harmonic introduces a possible corruption of the pure energy-per-mode concept assumed by MPFs.  In the scientific literature a `mode' is derived from the parameters used for the study of statistics which include the median, mean, and the mode.  In a field where Gaussian distributions abound thanks to the physics described by the Central Limit Theorem,\footnote{The extent to which the literature supports the Central Limit Theorem can be found in papers from 1937 through 2003 with Uspensky \cite{book:Uspensky}, Landon and Norton \cite{inproc:Landon}, Khinchin \cite{book:Khinchin}, North \cite{TR:North}, and Le Cam \cite{ppr:LeCam}.} the mode is merely the location on the abscissa, in terms of the frequency for a spectrum, at the maximum of the response.  In this sense the typical emphasis on energy-in-mode used by modal engineers is appropriate \cite{ppr:AllemangMAC03} to identify the deformed \emph{shape} of the eigenvector for that approximate center frequency.  However, transfer functions are inappropriate for these nonlinear systems, even if modified as Bendat does for simpler solitary nonlinear systems \cite{book:BendatNonlin98}.

\subsection{Contact nonlinear response}
\label{contactNonLin}


The author's CSC thesis \cite[App F]{MSthesis:kobold06} describes and plots the \emph{control law} as a sigmoid, using an arctangent function for the stiffness which is the slope of the load-deflection curve.  The control law is a mathematical representation of the nonlinear stiffness associated with the contact state between armor and hull, modeled by a sigmoid classifier.  This classifier is an arctangent function that defines the load-deflection curve, a smooth version of the typical bilinear gap element load-deflection table.  As the armor and hull make and break contact, the stiffness switches between high and low values of stiffness, respectively.  When the armor and hull are in contact, negative deflection (compressive penetration) results and there is a large stiffness due to the large slope of the deflection curve.  When the armor has separated from the hull, the nonlinear contact stiffness becomes a deflection curve of small slope.  This small value of extra stiffness provides computational stability.  The author's thesis \cite{MSthesis:kobold06} contains a description of the stability and application of the control law, demonstrating that the damping DOF and frequency DOFs are no longer related.  The following analysis expands on one of Dr. Winthrop's control laws listed in his December 2004 AFIT dissertation \cite{phdth:Winthrop04}, in a manner similar to how he analyzed a different control law.

%

Phase space (state space) plots \cite[Fig 12]{MSthesis:kobold06} validate its behavior.  The arctangent function that does the switching in this simple closed form model appears in state space formulation as show below in Equation \ref{singleDOFrelation}, which has the constants listed as Equation \ref{oneDOFconstants1}.  The state space description for the single mass speed, $v$, and acceleration, $ \dot{v} $, shown in Equation \ref{singleDOFrelation} is a function of the open and closed stiffnesses, $ k_{\mbox{open}} $ and $ k_{\mbox{\mbox{closed}}} $, for a damping coefficient of $ \zeta = d/m $ in units of hertz, where $m$ is a point mass.

\begin{equation} \label{singleDOFrelation}
\begin{array}{l}
  \dot{x} = v \\
  -\dot{v} = \frac{1}{2}\omega_o^2x(t)-\frac{1}{\pi}\tan^{-1}k_x x(t)
+ \frac{k_{\mbox{open}}}{k_{\mbox{closed}}} + \frac{d}{m}v \\
\end{array}
\end{equation}

This model of a ``welded-together" panel system\footnote{The term \emph{welded} assumes the plate deflections at both points are the same using an infinitely stiff connection. The closed gap stiffness used in this work is more like hard rubber \cite[App B.2.5]{MSthesis:kobold06}, but the effect is the same.} uses the parameters in Equation \ref{oneDOFconstants1}.  A ``manual," analytical calculation \cite[App C]{MSthesis:kobold06} helped validate the FEA results for the simulated vibration created before application of the laser vibrometry model.  Pepela's optically measured modes provided qualitative support for the application of the laser vibrometry model that used the FEA results to modulate the probe beam.  This section and the Appendix describes the former, a structural SDOF model.


\begin{equation} \label{oneDOFconstants1}
\omega_o = 2\pi 333.84, \frac{k_{\mbox{open}}}{k_{\mbox{closed}}} =
0.01, \frac{d}{m} = 0.02, k_x = 100  %
\end{equation}

In the arctangent gap model \cite[Fig 11]{MSthesis:kobold06} (not shown here), the stiffness transitions at contact $ x = 0 $.  Contact surfaces are imperfect due to microscopic protuberances that comprise the roughness of the surfaces.  As the two plates come into contact ($ x \leq 0 $) the roughness of the surfaces deform, compressing the protuberances, and a transition from low to high stiffness occurs rapidly in order to match the surrounding material. For a ``low" damping system, the damping is $\mu = 0.02$ kg/s and the open to closed dimensionless stiffness ratio is $k_{\mbox{open}}/k_{\mbox{closed}} = 0.01$.  The Appendix describes the dimensioned and dimensionless models.  A plot of speed versus gap opening in the thesis \cite[Fig 12]{MSthesis:kobold06} reveals that the high stiffness during compression of the base is a shallow orbit in $ x $ versus $ \dot{x} $ phase space. Changes in damping lead to changes in the range of the orbits during stabilization.  These control law formulations were implemented using a MATLAB system\cite{book:Polking99}.


Lyapunov function analysis shows the contact control law is asymptotically stable.  The Lyapunov function in this case is the total energy with simple damping loss.  In rare cases vibration energy gain to the hull upon which the armor plate clatters matches the cycle's damping energy losses and is in synch with plate contact. Persistent energy loss is due to structural damping, material heat losses due to bending, and fastener frictional losses.  The energy gain per cycle is less than the energy loss due to damping \cite[App F]{MSthesis:kobold06}. If the losses where small enough to just equal the gains, the system would be `stable in the sense of Lyapunov.'


This analysis above summarizes a clattering system.  Details of the physics of boundary conditions, initial conditions, and operating states is found in the thesis \cite[App F]{MSthesis:kobold06}.

\subsection{Laser vibrometry return from the probe beam}
\label{probedImaging}


The noise floor did not rise in Figure \ref{missedMode} but rather, the surrounding spatial areas had phase differences within the large spot size that, due to spatial averaging, underwent destructive interference in a bandwidth that removed most of the 1460 hertz mode.\footnote{Due to the location of Chladni lines of nodes for the 1460 Hz mode, the return contained Doppler shifts from different phases of the vibration for that mode.  Pepela's laser vibrometer \cite{MSthesis:NgoyaPepelaLaserVib03} used these Doppler shifts from different locations on the object that destructively interfered.  Other laser vibrometers measure speed with two pulses or use other techniques.}  This interference is a combination of optical phase shifts along-range caused by reflection from deflection amplitudes that are related to the \textit{structural} wavelength along the bar for each mode.

The purpose of the vibrating bar measurement was to show spectral elimination, but it also shows the discrete nature of high quality (low damping) structural modes.  The FEA in the CSC thesis \cite{MSthesis:kobold06} complemented this laboratory measurement \cite{MSthesis:NgoyaPepelaLaserVib03} by showing that, while spectral elimination can occur with structures that might be constructed in the lab (one dimensional modes shapes are the simplest super-symmetric structures), it is impractical to build commercial structures that are substantially super-symmetric.


The main result of the CSC simulations was that academic models can produce reductions of vibration sensitivity that theoretically verify part of Pepela’s spectral elimination thesis \cite{MSthesis:NgoyaPepelaLaserVib03}, within meaningful assumptions.  The single and two DOF models described here show why this is the case, theoretically.  A separate 2014 article \cite{ppr:insensitive} describes the lack of spectral elimination (or spectral reduction) for vehicle and other types of structures that are manufactured economically.

The lab vibration measurement shared some spectral reduction features seen in the nonlinear cross-spectral covariance simulation \cite{MSthesis:kobold06}.  The transient vibration results for the surface of the 3-D armor plate were the input for a MATLAB system that forms an image similar to one that a laser vibrometer would detect.  Commercial laser vibrometers typically only provide images processed with their proprietary system.  For the purpose of simulating the image of the vibrating plate, the CSC thesis used conventional Fresnel diffraction in a MATLAB system of functions.  The Fresnel propagation method introduced by Goodman \cite{book:GoodmanFO68} provides a computationally adequate method to propagate an image of the return from the target back to the detector.  In this sense, Pepela's laboratory measurement provided a validation for the FEA and vice versa.


The laboratory measurements in Figure \ref{allModes} and \ref{missedMode} use a clamped-clamped bar that has a high frequency fundamental mode (compared to vehicle fundamental modes that are below 50 Hz).  The spectrum has not dropped off with frequency by 3 kHz because this is still the ``lower frequency region" for the structure, as discussed in the section \ref{ZienkiewiczEnergyFlow}.  It has a stiff fixity meant to test the laser vibrometer for structural wavelengths that would undergo a difference in phase across the beam spot size.


The spectral elimination effect (Figure \ref{missedMode}) provided laser vibrometer manufacturers information on what spot size to suggest or program for their probe beams.  Laser vibrometry manufacturer's have wanted to avoid such `spectral elimination.'  The simulation and analysis in the CSC thesis \cite{MSthesis:kobold06} shows users and manufacturers that even specially manufactured structures are not super-symmetric and thus do not exhibit spectral elimination.


The 2006 CSC thesis \cite{MSthesis:kobold06} compares both structural and optically sensed CSCs, to imaging versus non-imaging returns.  The latter choice of observation state, non-imaging, is based on the analysis of a scalar signal composed of the spatial average of the entire image per time step.  Using only one scalar metric, non-imaging radiant flux simulations identify the modes on the target (HRA clattering on a hull) adequately, nearly as well as the imaging version.

\subsection{Finite element modeling and analysis}

\begin{figure}
\resizebox{8.4cm}{!}{
\includegraphics{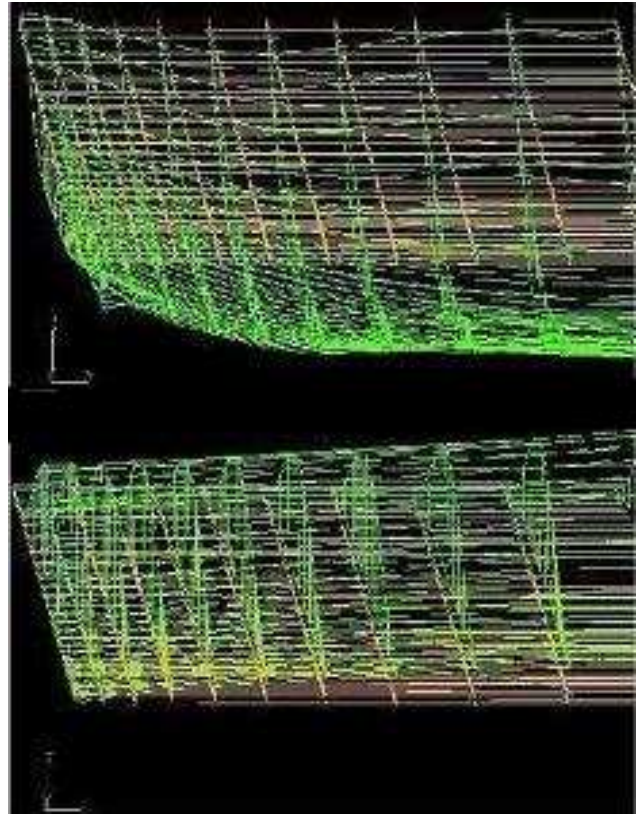}}
\caption[deformedShapeFEA]{\label{defShape} FEA of the deformed shape (green) for the clattering armor-on-hull model.  Exaggerated micron deflections would not be visible compared to the large physical geometry of the plates.  There is a 14.24 ms time difference (many time steps) between these two output time frames.  The views are \textit{into} the fore-shortened long direction (1 m) of the 0.5 m wide plate. The un-deformed shape is a light tan color.}
\end{figure}

The FEA produces a surface of vibration that is curved in 3-D and provides input for the scripts and functions run within MATLAB that produce an optical simulation of the target's image.  Unlike the small and large spot sizes in Pepela's measurement \cite{MSthesis:NgoyaPepelaLaserVib03}, the remote sensing of the armor-plate clattering uses a complete coverage large spot size such that all points on the armor modulate the probe beam.  Assume the surrounding clutter to be time-gated or otherwise removed.  The input to the MATLAB system is the set of displacements that simulate the transient dynamical response of the surface of the clattering armor for hundreds of time steps in a duration of a few seconds.  One such `vibration deformed shape' is seen in Figure \ref{defShape}.

\begin{figure}
\resizebox{8cm}{!}{
\includegraphics{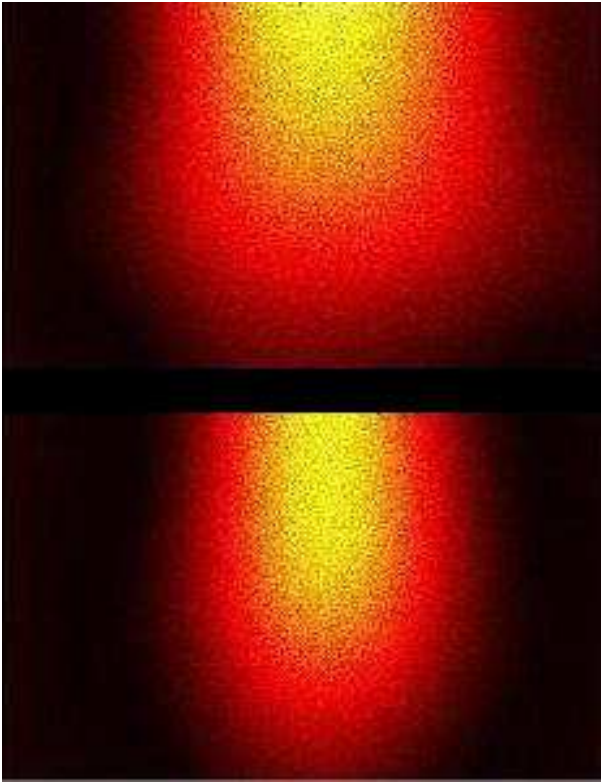}}
\caption[opticalImageHRA]{\label{imageOfReturn} Simulated magnitude of radiant flux image of the return after modulation by the vibrating plate in Figure \ref{defShape}.  The image horizontal width appears rotated 90 degrees from the long fore-shortened axis in the prior image in order to better display the 20 $\times$ 40 mm detector grid shape.  The maximum deflection ``hill" along the 1 m length shows up as a vertical maximum magnitude in the lower (later) pane of both figures.}
\end{figure}

The FEA representation of the vibration shape was applied to the model of the probe beam as a phase modulation.  Using Fresnel diffraction \cite{book:GoodmanFO68}, the return was imaged onto the detector 4 km away.  For the computers at that time, even when running MATLAB simultaneously on several machines on the Air Force Institute of Technology cluster, there was a delicate balance between adequate structural grid densities and optical grid densities for the spatial Fourier transform that performs the Fresnel propagation of the return.  A 10 $\mu$m  probe wavelength satisfied the numerical requirements for results shown in Figure \ref{imageOfReturn}.

A spectrum of the time sequence of sensed images, such as the two shown in Figure \ref{imageOfReturn}, appears in Figure \ref{impactResponseFRF}.  Detailed plots are in the 2006 CSC thesis \cite{MSthesis:kobold06}.

\section{Nonlinear eigen-states}
\label{nonLinEig}

Creating a simple model for a clattering structure may seem easy, until the fact that the eigenvalues must be nonlinear complex functions becomes apparent. While a two degree of freedom model developed in the Appendix herein was meant to validate the FEA, the FEA validates a major dynamical 2 DOF analysis result: Unsymmetrical modes increase in frequency for increases in contact stiffness. Investigation of these closed form simple damped sprung mass systems\footnote{`Sprung mass' systems are an automotive term for a systems where vibration is being isolated or suppressed.  At an academic or `free body diagram' level, they can sometimes be approximated by spring-mass-damper systems.} provides not only insight, but qualitative numerical validation. Whether an eigenvalue increases for anti–symmetric and un-symmetrical modes when gap stiffness increases provides another analytical basis for results seen in the FEA. Additionally, the behavior of the modes indicates that ‘symmetrical’ modes are better target identiﬁcation features than un-symmetrical modes. Hence the need to analyze the energy balance and stability \cite[App F]{MSthesis:kobold06} to validate the simple one and two DOF models as is summarized in the Appendix herein, including the time variation of amplitude and phase.


\begin{figure}
\resizebox{0.52\textwidth}{!}{%
  \includegraphics{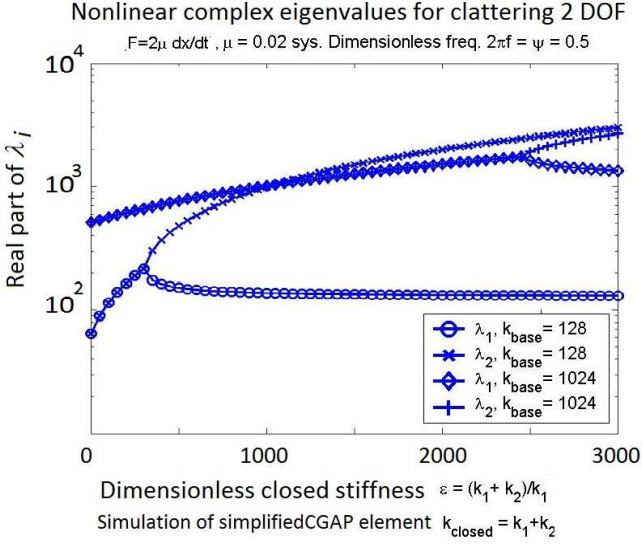}
}
\caption{\label{realEigen2dof} Real parts of the eigenvalues for the dimensionless DE show why higher-energy antisymmetric modes are less likely to be excited.  Above a $k_{\mbox{crit}}$ the symmetric (lower frequency branch) and antisymmetric (higher frequency branch) modes \cite[167]{book:Thomson88} for this two DOF problem break out into modes of well separated energy.}
\label{fig:1t-old}       
\end{figure}

Using the SDOF model as a basis, the Appendix uses Equation \ref{controLawEq} to provide the 2DOF solution.  Equation \ref{controLawEq} may appear deceptively simple because the time variation of the amplitude and phase is not apparent until explicitly formed.  Eigenvalues $\lambda_i$ are complicated functions of natural frequencies of individual modes which require `mode tracking' \cite{ppr:modeTracking95} with respect to stiffnesses \cite{phdth:modeTracking93}, due to their transient nature in reality  \cite{ppr:modeTrackingAnderson84}.  Figure \ref{realEigen2dof} plots the real part of the dimensionless eigenvalues of Equation \ref{controLawEq} as a function of dimensionless frequency versus dimensionless closed stiffness $k_{\mbox{closed}}$, scaled by the hull stiffness $k_1$ for a closed stiffness $\epsilon$ \cite{MSthesis:kobold06}.  Two models appear in the same plot.  One hull system is eight times stiffer than the other.

%

\begin{equation} \label{controLawEq}
m \frac{d^2x}{dt^2} + 2 \mu \frac{dx}{dt} + \omega_{sys}^2 \left( %
\frac{1}{2}+\frac{k_{\mbox{open}}}{k_{\mbox{closed}}}-\ %
\tan^{-1}\frac{100x}{\pi} \right) = 0 %
\end{equation}

From a chaos point of view, nonlinear attractors exist for the system defined by  Equation \ref{controLawEq}.  We know this because the systems exist in reality and their vibrations modes do fluctuate, albeit not monotonically.  The nonlinear attractors are related to the underlying hull forcing functions (D'Alambert's forces due to vehicle inertial loads), the timing and shape of the clattering, the plate stiffnesses, the structural and added damping (e.g., washers or armor-hull batting), and the fixity of the joints.

Initialy the two eigenvalues for the 2DOF system are equal.  As the closed stiffness increases a critical value of stiffness, $k_{\mbox{crit}}$, is reached where the armor can no longer follow the hull.  It starts to clatter in an \emph{unsymmetrical} mode.  (A 3-D contact surface cannot be anti-symmetric unless it is axisymmetric (1-D), but a 1-D models like the SDOF and 2DOF models can.)  When the stiffness exceeds $k_{\mbox{crit}}$ the two curves become distinct, branching into two separate paths.   This can be see for the plots of two models where the base stiffness is 128 and 1024 N/mm.  In the region before clattering, $\epsilon < k_{\mbox{crit}}$, the slope of eigenvalue curves, for the $8 \times k_{\mbox{closed}}$ higher base stiffness curve, decreases eight-fold compared to the smaller base stiffness.  At this higher 1024 N/mm base stiffness there is also an eight-fold increase in $k_{\mbox{crit}}$.  The intercept also increases 8-fold.  Derivations of these concepts, and further results, including how the imaginary parts of the frequencies (not shown in Figure \ref{realEigen2dof}) contribute to the transfer of energy between the modes are developed in the Appendix.


\section{Analysis and conclusion}
\label{analysisConclusion}

A structural vibration system model of a plate contact system was shown to have effectively discrete low frequency modes.  Both SDOF and 2DOF models derived in the Appendix have qualitatively verified the symmetrical versus unsymmetrical modes, and the higher frequency of the unsymmetrical clattering modes.  The full finite element model transient results also showed the tendency for vibration strain-energy to collect in the lower frequency modes, as model engineers would attest, and as was calculated analytically in the classical literature \cite[340, 341]{book:Zienkiewicz91}.

\begin{figure*}
\centerline{Table \ref{TableI}: Comparison of Observational Characteristics}
\begin{tabular}{|r||l||l||l|}
\hline
\label{TableI}  Characteristic & 1. Discrete modes & 2. Energy transitions & 3. Energy ordered by mode \\
\hline
\hline
 Concept sketches & Figure \ref{discreteModeSketch} ($\Sigma_j^N E_j$) & Figure \ref{energyXfrSketch} ($ E_j \rightleftharpoons E_{j+1} $) & Fig. \ref{modalOrderSketch} ($ E_j \Longleftarrow E_{j+1} $) \\
\hline
 Dependency & High Q resonances & Clattering FEA model & Same deflection $\propto \omega^2$ at high \textit{f} requires $E_j \nearrow$ \\
\hline
 Spectral effect & Figure \ref{allModes} and \ref{missedMode} & Figure \ref{impactResponseFRF} & Damping, $\lim_{f\nearrow}c(f) \propto f$ eschews $E_{strain}$  \cite{book:Zienkiewicz91} \\
\hline
 Character & Low \textit{f} modes & Nonlinear joints & Steady state vibration state $E_i \propto 1/f$ \\
\hline
\hline
 Small spot size & Sees all modes & Blind to many $\Delta E_{jk}$ & No Spectral elimination (SE) \\
\hline
\hline
 Large probe spot & SE in bars & Sees more $\Delta E_{jk}$ & Some SE (spectral reduction) \\
\hline
 Super-symmetric & Some missed modes & Misses most $ E_j \rightleftharpoons E_{j+1} $ & Provides th. of SE \cite{ppr:insensitive}. Target rarely super-sym. \\
\hline
 Manufactured parts & Sees all modes, Fig \ref{impactResponseFRF} & FRF, coh(\textit{f}), MPFs & Typ. Figure \ref{impactResponseFRF} (super-sym. target is \emph{rare}) \\
\hline
\end{tabular}
\end{figure*}


(1) The lab tests on the doubly clamped bar show experimentally that high SNR modes are effectively discrete.  In the absence of substantial damping they are not just the maxima of a spectrum, but spikes in the spectral response \cite{MSthesis:NgoyaPepelaLaserVib03}.  The system in Figure \ref{allModes} has a signal-to-noise-ratio that is huge.  These massive SNRs show that the lower modes of structure, that are fairly high in frequency for this case due to the double clamped nature of the bar, can easily be considered discrete.  Hence Q.E.D., the observational characteristic (1) is demonstrated.

Exposition of nonlinear eigenvalues provides results for theory and simulation for common nonlinear structures.  Symmetrically moving parallel plates have more strain (and thus more strain-energy) than clattering plates where the unsymmetrical motion interrupts their nearly sinusoidal in time out-of-plane-motion to spew strain-energy into acoustics and even permanent set (deformation of the material).  Through simulation\footnote{Figure \ref{impactResponseFRF} shows that decreasing baffle stiffness between the plates (see its legend) drives unsymmetrical resonances lower.} and by experience the observational characteristic (1) appears true, that lower modes are effectively discrete for high quality systems (low damping), and that, (2) and (3), damping and clattering help the energy flow into the lower modes from higher frequency modes, when there are pathways (nonlinear joints) that allow energy transfer -- although overall energy is lower as damping increases (see Figure \ref{impactResponseFRF}).  Most vehicle components have bolted or riveted joints that allow such energy transfer \cite{book:BendatNonlin98} as can be seen with plots of coherence spectra, coh(\textit{f}) \cite{book:BendatPiersol93}.

(2) The tools in industry that provide MPFs are mature engineering methods from the 1990's such as the FEA tools put in place, for example, by MacNeal Schwendler Coporation engineers in NASTRAN\textsuperscript{\textsc{tm}} \cite{inProc:MPFbyTedRose98}.  These MPFs stored in the FEA vector $\Phi$ can measure the energy transmitted between states (between eigenvectors of the system) where $\Delta\Phi_t(f_i) = \Phi_{t+1}(f_i) - \Phi_t(f_i)$ for mode \textit{i}.

(3) Modal engineering (i.e., ride-and-handling) is a trade secret endeavor (except for the Pininfarina publication \cite{inProc:Pininfarina}).  It makes use of analysis tools in order to calculate modal participation factors.  These MPFs show the `participation' (energy per mode) using modes developed from normal modes output of FEA eigenvectors.

Therefore, due to the (3) energy ordering of modal states for complicated system synthesis models and the ordinary vehicles they represent, the work of vehicle design for these issues focuses on (2) transition probabilities\footnote{Transition probabilities are related to CSCs calculated in the CSC thesis \cite{MSthesis:kobold06} and the coherence spectrum \cite{book:BendatPiersol93}.} and (1) energy levels $\Phi(f_i)$.  It would be useful to see calculations of other types of oscillator damping that produce an ordering of energy levels similar to that described by Zienkiewicz \cite[340-341]{book:Zienkiewicz91}.  These concepts are summarized in Table \ref{TableI}.

Pencil-thin probe beams like that used in the lab measurement have observational characteristics quite different from the large spot size used for the FEA-MATLAB model of fully illuminated clattering armor.  The former is less susceptible to spectral elimination but the latter is adequate for spectral identification of economically manufactured vehicles.  Such full coverage probe beams are less likely to be subject to spatial coherence issues or illuminate a node of the Chladni zone \cite{book:ChladniWiki} that can cause pencil-thin beam returns to fail \cite{book:RayleighTOS1}, since for the large (full coverage) spot size, some of the beam is still getting through.

\section{Appendix -- Two-DOF contact eigenvalues}
\label{nonlinearEigenvaluesAppendix}

This Appendix summarizes the nonlinear dynamics of a simplified one-dimensional (1-D) form of a structural contact system, and its relationship to the full 3-D FEA.

\subsection{Closed-form SDOF nonlinear contact response}
\label{closedFormSDOF}

This subsection displays a copy of the MathCad\textsuperscript{\textsc{tm}} output for the closed form solution to the damped single DOF (SDOF) oscillator meant to represent a lumped mass model of the homogeneous rolled armor (HRA) to hull vibration. The mechanics are based on the nonlinear solutions in Dr. Major Winthrop's dissertation with a simplification of the control law (shown in the next subsection) that models simple contact. $x(t)$ of Equation \ref{x1DOFsolution} is Dr. Winthrop's relation \cite[Eq 3.14]{phdth:Winthrop04}.


The SDOF calculations shown below, with suitable simplifications and assumptions, are used in the next subsection for the 2DOF solution. This SDOF description describes the effect of contact stiffness (bi-linear stiffness nonlinearity) on a solution
composed of symmetric and antisymmetric one dimensional modes.  This solution applies to the symmetrical and unsymmetrical modes (respectively) resulting from the FE model for the tank hull and armor.  The SDOF model is a severely lumped-mass model. Different symmetry adjectives are necessary for the three dimensional FE model because clattering modes where the armor is moving opposite to the hull does not have the laterally spatial uniformity implied by the word `antisymmetric.'  Hence, for 3-D systems the latter symmetry adjectives are appropriate whereas for the 1-D system the former simpler words apply.

At first the derivatives act on the dimensionless system assuming all variables are nonlinear.  After starting with a restricted case, we can bring the variables into explicit nonlinear use, one at a time, to refine the calculation.

The state-space representation \cite{book:ChenSlinSys} is shown here in its phase space form, ($x(t)$ and its derivatives).\footnote{The state-space definition of the system combines the state $x(t)$ with its output $y(t)$ and changes in state ($\partial x/\partial t$).  These matrix relationships \cite{book:Brogan} of the state change vector ($\dot{x}$) and system output and its state and input ($u(t)$) are one of many sets of $A, B, C$, and $D$ matrices, easily confused with other field's ABCD systems such as optical ABCD ray matrices for laser resonator orientation \cite{book:laserSilfast,book:lasersSvelto98}.
\begin{list}{}
  \item $\dot{x} = Ax + Bu$  using the state and input matrices
  \item $y = Cx + Du$ via output and feed-through matrices
\end{list}
All matrices and variables can be functions of time.  Some definitions distinguish state space as discrete compared to continuous phase space, However the field of linear systems often uses continuous output and even state variables and the resulting Kalman filter is nearly ubiquitous \cite{book:ChenSlinSys}.}  With the definition $ \qquad \cos \phi_{\psi\beta} \equiv \cos \left[ (\dot{\psi} t + \psi_o) t + (\dot{\beta} t + \beta_o) \right] $,

\begin{equation} \label{x1DOFsolution}
x(t) = (\dot{a}(t) t + a_o(t)) e^{-(\dot{\mu} t + \mu_o)t}  %
\cos \phi_{\psi\beta}
\end{equation}

Even the first derivative, Equation \ref{x1DOFsolution2}, is non-trivial.

\begin{equation} \label{x1DOFsolution2}
\dot{x}(t) = \frac{d}{dt} \left[ %
(\dot{a} t + a_o) e^{-(\dot{\mu}(t) t + \mu_o(t))t} %
\cos \phi_{\psi\beta} \right]
\end{equation}

%


These relations will look simpler after some assumptions are worked out below.  This output from MathCad allows factoring in several ways.  Multiplying by the inverse of the damping, then collecting ($\dot{x}(t) e^{(\dot{\mu} t + \mu_o) t}$) on $ \cos \phi $ and then on $ \sin \phi $, Equation \ref{xdotDOFexpression1} provides for a somewhat compact expression of the speed.

\begin{equation} \label{xdotDOFexpression1}
\begin{array}{l}
  \mbox{The ``undamped" speed} \qquad
\dot{x}(t) e^{(\dot{\mu} t + \mu_o) t} = \\
  \left( %
2 \dot{a} \dot{\mu} t^2 + ( 2 a_o \dot{\mu} + \dot{a} \mu_o) t +
 a_o \mu_o -\dot{a} \right) %
 \\
\times \cos \left( %
\dot{\psi} t^2 + (\psi_o + \dot{\beta}) t + \beta_o) \right) %
 \\
  - \left( %
(2 \dot{a} \dot{\psi} t^2 + ( 2 a_o \dot{\psi} + \dot{a} \psi_o +
\dot{a} \dot{\beta}) t
+ a_o \psi_o - a_o \dot{\beta} \right) %
\\
\times \sin \left( %
\dot{\psi} t^2 + (\psi_o + \dot{\beta}) t + \beta_o \right) %
 \\
\end{array}
\end{equation}

With more assumptions restricting the nonlinearity of the solution, the acceleration is also collected on $ \cos \phi $ and then $ \sin \phi $ in Equation \ref{xddotDOFsolution} for this simplified expression:

\begin{equation} \label{xddotDOFsolution}
\begin{array}{r}
  \ddot{x} = \left( %
 - \dot{a} \psi_o t + [\dot{a} \mu_o - a_o \psi_o^2] \right) %
\cos (\psi_o t + \beta_o)
\\
-(\dot{a} \mu_o \psi_o t + a_o \mu_o
\psi_o ) \sin (\psi_o t + \beta_o)
  \\
  \forall \qquad \dot{\mu} = 0, \quad  \dot{\beta} = 0, \quad \dot{\psi} = 0 \\
\end{array}
\end{equation}

The first order nonlinear solution in Equation \ref{xddotDOFsolution} uses a constant amplitude \textit{a}.  While not explicitly a function of time, it has a constant time rate of change, $ \dot{a} $.  A further order of nonlinearity to allow the amplitude change rate to be a function of time would follow in priority the use of $ \dot{\psi} $, the rate of change of a dimensionless frequency. Further refinement is left to the future.  The following subsections describe how Figure \ref{realEigen2dof} shows that while the stabilizing stiffness increases above the critical lift-off frequency, symmetrical modes remain at a constant frequency while un-symmetrical modes increase in frequency.

\subsection{Closed-form MathCad 2 DOF contact}
\label{closedFormMathCad2dof}

The subsections to follow display MathCad symbolic output for the two DOF (2DOF) model.  They are the closed form solutions for a damped 2DOF oscillator.  $k_x$ is the ``rate" that defines the sharpness of the contact.


\subsection{Two DOF DE's and solutions}

Equation \ref{DEfor2DOF} is a dimensioned form of the two DOF damped oscillator DE where $P$, $m$, $k$, \textit{d}, $\epsilon$, $\xi$, and \textit{t} are applied force, outboard mass, foundation stiffness, damping, control law stiffness, surface displacement, and time.

\begin{equation} \label{DEfor2DOF}
\begin{array}{l}
  \left(%
\begin{array}{c}
  P_1(t) \\
  0 \\
\end{array}%
\right) = \left(%
\begin{array}{cc}
  m_1 & 0 \\
  0 & m_2 \\
\end{array}%
\right) \frac{d^2}{dt^2} \left(%
\begin{array}{c}
  \xi_1 \\
  \xi_2 \\
\end{array}%
\right) + %
\\
 \left(%
\begin{array}{cc}
  d_1 & 0 \\
  0 & d_2 \\
\end{array}%
\right) \frac{d}{dt} \left(%
\begin{array}{c}
  \xi_1 \\
  \xi_2 \\
\end{array}%
\right) + \\
  \Bigg[ \left(%
\begin{array}{cc}
  1 & 1 \\
  1 & 1 \\
\end{array}%
\right) + \left(%
\begin{array}{cc}
  \epsilon_1 & 0 \\
  0 & \epsilon_2 \\
\end{array}%
\right) \bigg(\frac{1}{2} - \frac{1}{\pi} \left(%
\begin{array}{cc}
  \arctan(r \xi_1) & 0 \\
  0 & 0 \\
\end{array}%
\right) \bigg) \Bigg]%
\\
\times \left(%
\begin{array}{cc}
  k_1 & -k_1 \\
  -k_1 & k_1 + k_2 \\
\end{array}%
\right) \left(%
\begin{array}{c}
  \xi_1 \\
  \xi_2 \\
\end{array}%
\right) \\
\end{array}
\end{equation}

\textit{k} will be the dimensionless stiffness ratio for the second oscillator, $ k = (k_1+k_2) / k_1$ which is the stiffness of the base of the system.  $ \xi $ indicates the dimensioned form of location in the dimensionless \textit{x} direction.  The dimensioned control law for structural contact for this
two DOF problem is the displacement $ u(\xi) $ of Equation \ref{2DOFcntrLaw}.

\begin{equation} \label{2DOFcntrLaw}
u(\overrightarrow{\xi}) = \Bigg[ \frac{1}{2} - \frac{1}{\pi} \left(%
\begin{array}{cc}
  \arctan(k_x \xi_1) & 0 \\
  0 & \arctan(k_x \xi_2) \\
\end{array}%
\right) \Bigg]
\end{equation}

Use of Winthrop's method \cite{phdth:Winthrop04} results in a dimensionless form of Equation \ref{dimlessDEfor2DOF}.

\begin{equation} \label{dimlessDEfor2DOF}
\left(%
\begin{array}{c}
  F_{\mbox{applied}}(t) \\
  0 \\
\end{array}%
\right) =
\end{equation}

$
\left(%
\begin{array}{cc}
  1 & 0 \\
  0 & 1 \\
\end{array}%
\right) \ddot{\overrightarrow{x}} + 2 \left(%
\begin{array}{cc}
  \mu_1 & 0 \\
  0 & \mu_2 \\
\end{array}%
\right) \dot{\overrightarrow{x}}%
 +  \left(%
\begin{array}{cc}
  1 + \epsilon_1 u(x_1) & -1 \\
  -1 & k \\
\end{array}%
\right) \overrightarrow{x} %
$


This relation uses the vector $ \overrightarrow{x} = [x_1 \quad x_2]^T $, the dimensionless location of the two masses.  Continuing with assumptions outlined in the Winthrop dissertation \cite{phdth:Winthrop04}, assume a straightforward solution. Equation \ref{assumedSolutionX} shows all the variables that vary with time.


\begin{equation} \label{assumedSolutionX}
x_i = a_i(t) e^{\mu_i(t) t} \cos(\psi_i(t) t + \beta_i(t))
\end{equation}

Substituting assumed solutions from Equation
\ref{assumedSolutionX} into Equation \ref{dimlessDEfor2DOF}
results in a Special Eigenvalue Problem (SEVP).  The dimensionless system frequencies $ \psi_i $ are functions of the individual frequencies $ f = \omega /2\pi $ from
$ \omega_i^2 = (k_i / m_i) - 2\zeta_i^2 $ where $ \zeta_i =
d_i / 2m_i $.  (See the details in  \cite{phdth:Winthrop04}.)  In order to assist in making the solutions tractable, exclusion of some forms of time variation and
non-uniformity is appropriate.  The first step of approximation is to
restrict time variation of the amplitude, damping, and phase, and to
remove the driving load.  Equation \ref{LinearityAssumptions2DOF}
lists these assumptions of linearities and uniformity that
restrict us to a `first order' nonlinearity, a time variation of
the dimensionless frequency, $ \psi $, which is appropriate for the common task of mode tracking \cite{ppr:modeTrackingAnderson84,phdth:modeTracking93}.


\begin{equation} \label{LinearityAssumptions2DOF}
\begin{array}{r@{\quad}}
  a_1(t) = a_2(t) = constant = a \\
  \mu_1(t) = \mu_2(t) = constant = \mu \\
  \beta_1(t) = \beta_2(t) = constant = \beta \\
  F_{\mbox{applied}}(t) = 0 \\
\end{array}
\end{equation}


In words, these simplifications are:

\begin{itemize}
  \item Constant amplitude `a' cancels out of the DE.
  \item Uniform damping $ \mu $ is for simplicity.
  \item But uniform phase $\beta$ is realistic.
  \item And finally, solve for free vibration first.
\end{itemize}

Using a new variable for dimensionless phase, $ \phi = \psi t + \beta $, to simplify the state space system (Equations \ref{x1DOFsolution} and \ref{x1DOFsolution2}), the solution starts with a definition of the speed (Equation \ref{FirstOrder2DOFspeed}) and acceleration (Equation
\ref{FirstOrder2DOFaccelDef}).

\begin{equation} \label{FirstOrder2DOFspeed}
\dot{x} = -\Big[ a\mu e^{-\mu t} \cos \phi + a\mu e^{-\mu t} \sin
\phi  \Big]
\end{equation}

Maintain the coefficients as variables of time in the subsequent derivative:

\begin{equation} \label{FirstOrder2DOFaccelDef}
\ddot{x} =  a (\mu^2 - \psi^2) e^{-\mu t} \cos \phi + a (\mu \psi
+ \mu \psi) e^{-\mu t} \sin \phi
\end{equation}

Using the format of Equation \ref{dimlessDEfor2DOF}, the resulting dimensionless DE in Equation \ref{FirstOrder2DOFdimless2} below starts to look like a SEVP (Equation \ref{FirstOrder2DOFmatrixEq}), easily solved by eigenvalue methods.  We move representation of the time derivatives into the coefficients of the vectors in operator form so that the vector $\overrightarrow{x} = [x_1 \quad x_2]^T $ factors out.  Here we carry the decay (extinction) constant and frequency, $ \mu $ and $ \psi $ respectively, until getting the formal solution (``nonlinear" eigenvalues and eigenvectors).  A purely imaginary frequency $ \exp(if't) =  \exp(i[i\mu]t) $ is an exponential term in $ \exp(-\mu t) $.  In this case the sign of $ \mu $ indicates damping.  The ensuing 1-D math and descriptions provide insight into the dynamics and validate the transient `nonlinear' modes output by the 3-D FEA modes that exist in reality.

\begin{equation} \label{FirstOrder2DOFdimless2}
\begin{array}{l}
  \left(%
\begin{array}{cc}
 \Pi_1  & 0 \\
 0 & \Pi_2 \\
\end{array}%
\right) \left(%
\begin{array}{c}
  x_1 \\
  x_2 \\
\end{array}%
\right)
 + \\
  \left(%
\begin{array}{cc}
\Omega_1  & 0 \\
 0 & \Omega_2 \\
\end{array}%
\right) \left(%
\begin{array}{c}
  x_1 \\
  x_2 \\
\end{array}%
\right) + \\
  + \left(%
\begin{array}{cc}
  1 + \epsilon_1 u(x_1) & -1 \\
  -1 & k \\
\end{array}%
\right) \left(%
\begin{array}{c}
  x_1 \\
  x_2 \\
\end{array}%
\right) = \left(%
\begin{array}{c}
  0 \\
  0 \\
\end{array}%
\right) \\
\end{array}
\end{equation}

The time derivatives come from application of differential operators using Equations \ref{FirstOrder2DOFspeed} and \ref{FirstOrder2DOFaccelDef}:

$ \qquad \qquad \Pi_i \equiv (\mu_i^2 - \psi_i^2) + 2\mu_i\psi_i\arctan \phi_i, $

$ \qquad \qquad \Omega_i \equiv 2\mu_i \bigg[-\mu_i - \psi_i\arctan \phi_i \bigg]$

These relations use a different dimensionless stiffness, $ \epsilon \equiv \epsilon_1 = k_{\mbox{closed}} / k_{\mbox{open}} $. Equation \ref{FirstOrder2DOFdimless2} is of the SEVP format as shown below in
Equation \ref{FirstOrder2DOFmatrixEq}.

\begin{equation} \label{FirstOrder2DOFmatrixEq}
\left(%
\begin{array}{cc}
  A_{1,1} & A_{1,2} \\
  A_{2,1} & A_{2,2} \\
\end{array}%
\right) \left(%
\begin{array}{c}
  x_1 \\
  x_2 \\
\end{array}%
\right) = \left(%
\begin{array}{c}
  0 \\
  0 \\
\end{array}%
\right)
\end{equation}

Equation \ref{FirstOrder2DOFmatrixAdef} shows the format of the
system matrix A for submission to an eigen-solver.  Some of the
terms in $ A_{1,1} $ and $ A_{2,2} $ were kind enough to cancel.

\begin{equation} \label{FirstOrder2DOFmatrixAdef}
\mathbf{A} = \left(%
\begin{array}{cc}
  A_{1,1} & A_{1,2} \\
  A_{2,1} & A_{2,2} \\
\end{array}%
\right) = %
\end{equation}

$
\qquad
\left(%
\begin{array}{cc}
  -\mu_1^2 - \psi_1^2 + 1 + \epsilon u(x_1) & -1 \\
  -1 & -\mu_2^2 - \psi_2^2 + k \\
\end{array}%
\right)
$

Calculation of the determinant of \textit{A} with the 1997 version of MathCad\textsuperscript{\textsc{tm}} provides solutions to the characteristic polynomial of the system.\footnote{MathCad '97 has a small Maple kernel that is far easier to use than recent 2016 versions - A 15 minute great solution versus days of dealing with customer support and still not quite getting the better solution.}  So there are two sets of solutions: free classical vibration, and the vibration solution where the parts are welded together (modelled as ``welded" is equivalent to epoxied with hard rubber in the FE model \cite{MSthesis:kobold06}).  The latter reduces to the former for slight contact, $u$ = 0, before compression ($u < 0$), as defined in Equation \ref{2DOFcntrLaw}.  Therefore, the ``welded" solution gives an indication of the dynamics by investigating the behavior of variations in control law $u$ from zero to one. The switching `control law' for the contact, $u(x)$, helps determine the actual frequencies. These frequencies have a cyclic variation, one value for open gap and another for closed.  But as the nonlinear FEA results show, both systems are active on average.  Energy flows into and out of both `open gap' and `closed gap' modes when averaged over many cycles.  The contact gap system is a time composite system.  Its dynamics are best understood with modal analysis techniques. The noise and vibration industry developed these techniques with nonlinearities such as contact in mind \cite{phdth:Allemang80,inbook:HarrisVib96,inbook:HarrisModal96,inbook:HarrisShock96}.

The energy flow oscillation, as the system states evolve in an oscillatory manner, might be a useful model for other systems where the physics may appear to forbid energy level ordering.  Such disorder may happen on a time frame we are as yet unaware of.  However, for vehicle structures, experience indicates that the oscillation period is usually on the order of a minute to an hour.

The eigenvalues in Equation \ref{FirstOrder2DOFeigenvalues} are therefore a dual set for zero and nonzero control law values, $u(x)$ for an open and closed gap, respectively.  In the dimensionless system, the non-contact state has unity stiffness in DOF 1 and stiffness of k for DOF 2.  But DOF 1 adds to the stiffness when in contact, so that base plus contact stiffness becomes $ 1 + \epsilon $.  This is the FEA maxim that `stiffnesses add" \cite{misc:aero51xAnderson93}.  The derivation of the dimensionless ``eigenvalues" $\lambda_i$ in Equation \ref{FirstOrder2DOFeigenvalues} came from the Mathematica results discussed further in the CSC thesis \cite{MSthesis:kobold06}.

\begin{equation} \label{FirstOrder2DOFeigenvalues}
\left(%
\begin{array}{c}
  \lambda_1 \\
  \lambda_2 \\
\end{array}%
\right) = %
\\
 \frac{1}{2} \left(%
\begin{array}{c}
  (1 + \epsilon + k)-2(\mu^2 + \psi^2) - \sqrt{\kappa} \\
  (1 + \epsilon + k)-2(\mu^2 + \psi^2) + \sqrt{\kappa} \\
\end{array}%
\right)
\\
\end{equation}

$ \kappa \equiv 5 - 2 \epsilon + \epsilon^2 + 2 k - 2 \epsilon k - k^2 $

For subsequent nonlinear calculations we assume $u = 1$ so that contact is active.  Assume $ \epsilon > k $ (dimensionless) and for stability $ \epsilon_1 > k_2 = k_{\mbox{base}} $ (dimensioned).

The real part of values of the eigenvalues are the data for the plots in Figure \ref{realEigen2dof}.  The plots makes it clear that after a critical $\epsilon$, the higher frequency mode (the clattering, antisymmetric mode) separates from the symmetric mode.  When the hull stiffness is far enough away from the critical stiffness, the symmetric mode settles into a constant eigenvalue with respect to $\epsilon$ regardless of the dimensionless stiffness.  ($\partial\lambda_1/\partial\epsilon$ and $\partial\lambda_1/\partial k \rightarrow 0$ but not for $\lambda_2$.)  The higher frequency mode is the clattering mode which from the 3-D FEA results in Figure \ref{impactResponseFRF}, and by experience and by Zienkiewicz \cite[340-241]{book:Zienkiewicz91}, is clearly the lower energy mode.  It takes far more energy to excite an antisymmetric mode to have the same deflection amplitude as a symmetric mode.  Provided an avenue (joint or structural connection) to excite the lower mode, the energy goes into the lower mode - as any child whipping a taut cord to try to vibrate it at higher modes comes to understand for a moment.


\subsection{Two DOF nonlinear contact eigenvalues}
\label{closedForm2dofNonLin}

Application of the SDOF model over all DOF's using a theory of linear structural response gives a relationship for a linear transfer function \cite{book:BendatPiersol93} (the Fourier transform of the impulse response \cite{book:ChenSlinSys}).  $ K_i $ is the generalized stiffness for DOF number \textit{i}.  Each of the many DOFs in a linear time-invariant (LTI) system has a relationship that has the same form as shown in a transfer function \cite[p. 81]{MSthesis:kobold06}.  But these functions are often inapplicable.  Common vehicles are filled with nonlinear joints, all of which can be adequately modeled in FEA with some effort. For the ensuing 2 DOF system, the results of the prior subsections and the 2 DOF extension of the SDOF theory are formulated for entry into Mathematica.  This subsection shows the reformulation and the
Mathematica results.


\subsection{Eigenvalues: Damped 2 DOF sprung mass}

Repeated roots in the eigenvalues $\lambda(\epsilon, k)$ plotted in Figure \ref{realEigen2dof} in Section \ref{nonLinEig} occur for stiffness lower than $ k_{crit} $.  They indicate the symmetric and antisymmetric modes have the same frequency but that there are real and imaginary parts to the eigenvalues \cite[171]{book:Thomson88}. This implies that there is a growth in energy for one mode and a decay in the other.  During the cycling of loads on most joints, the contact frictional footprint area also oscillates, which varies the stiffness.  Therefore, the imaginary parts of the frequencies act to re-balance the strain-energy.  The control law $u(t)$ approximates this physics.  Over time energy will move from the higher energy antisymmetric mode to the lower frequency symmetric mode, unless it is suppressed.

However, the low base stiffness situation (repeated roots) merely indicates that the stiffness of the base is so low that the close and far masses (a one -- dimensional lumped mass model for the hull and armor) vibrate together as if in free space.  Either they start with negligible relative motion where both together vibrate away from and towards the base (the trivial solution for low base stiffness) where the center of mass is oscillating according to the small base stiffness ($ f_{\mbox{CG}}^2 = k_{\mbox{base}} / 4\pi^2[m_{\mbox{hull}} + m_{\mbox{HRA}}] $),  or they vibrate apart and together with their center of mass remaining stationary.  The eigenvalue increase for anti-symmetric modes when gap stiffness increases provides a basis for similar stiffness-related results seen in the unsymmetrical FEA modes. Additionally, \textbf{this nonlinear eigenvalue behavior indicates that certain modes (`symmetrical') are better target identification features than others} if the deflection at higher frequencies is too small for adequate modulation of the probe beam.

\subsection{1-D eigenvectors compared to 3-D FEA}


This DE (Equation \ref{FirstOrder2DOFdimless2}) and the SEVP (Equation \ref{FirstOrder2DOFmatrixEq}) will be shown to be the `special' eigenvalue problem system matrix $ [A_1] $ for a simplified nonlinear contact system.  The system matrix $ A_1 $ provides the special EVP where $ [A_1] \overrightarrow{\Lambda}_i = zero $ for each eigenvector $ \overrightarrow{\Lambda_i} $ in the matrix of (two) eigenvectors $ \mathbf{\Lambda} $.  (This is still a 1-D system, but with 2 DOF, hull deflection and armor deflection.)  There are three terms in the dynamical D.E. (such as Equation \ref{dimlessDEfor2DOF}) but we can see that the system matrix is [A] from $ [A_1] = |[A]-\lambda [I]| $. The simplifications include that the damping is uniform for both the base and contact stiffness line elements; damping from the fixed base to the concentrated point mass modelling the hull is the same coefficient of the speed as the damping for the relative motion of the armor versus the hull.  The combination of low base stiffnesses and high switching stiffness (due to use of a stabilizing ``open" stiffness-) is stable.  (Not shown here, see CSC thesis \cite{MSthesis:kobold06}.)

The more severe assumption is that dimensionless frequencies are equal, $\psi_1 = \psi_2 $ (setting the \textit{natural} frequency of both the hull and the armor to be the same). While this is a strong assumption, it was understood that the FEA will give proper simulation results for all modes. Also, with this statement, we are saying that the particular solution is the same for both DOF's which is not actually the case, except for purely symmetric and antisymmetric modes.  However, we know that physically symmetric and antisymmetric modes, specifically these two modes alone, comprise the complete set of time solutions in Equation \ref{nonlinearSolution} for the 2DOF undamped SEVP considered. Hence, the assumption $\psi_1 = \psi_2 $ gets validation from the physical dynamics of this two DOF undamped system.  Damping will add a complication to this system but to first order assume both masses have the same particular order of solution, set for each time step.

\begin{equation} \label{nonlinearSolution}
x(t) = \overline{\frac{x_{\mbox{max}}}{L^\ast}} \bigg(\cdots\mbox{nonlinear
terms}\cdots\bigg) e^{j\psi t + \phi_o} %
\end{equation}

$ \qquad = a(t)e^{-\mu_i t}\cos(\phi_i t
+ \beta(t)) $

\begin{equation} \label{systeMatrix2DOF}
[A_1] =  %
\\
\left(%
\begin{array}{cc}
  -\mu^2 - \psi^2 + 1 + \epsilon & -1 \\
  -1 & -\mu^2 - \psi^2 + 1 + k_{\mbox{open}} \\
\end{array}%
\right)
\end{equation}

\textbf{The dimensionless frequencies, $\psi $, just happened to be in this system matrix in the same form as an eigenvalue in the form $ A_1 = |[A] - \psi^2 [I]| $.  This form where $\psi_1^2 =\psi_2^2 = \lambda_i $  are the eigenvalues, only applies to the assumptions of equal damping and frequency for the two masses.} Usually there is a superposition of both symmetric and antisymmetric modes so the hull -- HRA subsystem will resonate at both $ \psi_1 = \pm \sqrt{\lambda_1} $ and $ \psi_2 =
\pm \sqrt{\lambda_2} $ in a linear combination of modes.  Each of the two $ \Lambda_i $ modes will have both $\psi_1$ and $\psi_2$ active for that one mode. Therefore, the energy applicable to both $\psi_1$ and $\psi_2$ for $ \lambda_1 $ is the same.  This energy will generally be different from that for both $\psi_1$ and $\psi_2$ for $ \lambda_2 $.

The FEA eigenvalues (the values of the diagonal NASTRAN $ \Lambda $ matrix) are related to their eigenvectors $ \overrightarrow{u_i}^T $, as plotted in the thesis \cite[Fig21-23]{MSthesis:kobold06}. The square roots of the eigenvalues, the FEA modal frequencies, are summarized in the same thesis \cite[Table 7, Fig 20]{MSthesis:kobold06}. Those frequencies and plots of $ \overrightarrow{u_i}^T $ represent eigenvalues {$ \lambda_i $} within $ \Lambda $ and eigenvectors $ \Gamma_i $ within $ \Sigma $, in analogy to the 2DOF model. As discussed above, the similar-frequency argument makes physical sense.  Yet, the similar damping assumption does not have a physical rationale; it is an approximation.  It is necessary to keep this calculation simple enough for the extra $\psi$ terms to cancel.  (The FEA and experience validate this technique for use in this case.)

\begin{equation} \label{MathematicaDE}
A_1 = \left(%
\begin{array}{cc}
  1 + \epsilon -\mu^2 - \psi^2 & -1 \\
  -1 & k -\mu^2 - \psi^2 \\
\end{array}%
\right)
\end{equation}

The eigenvalues in matrix form are $\Lambda = [ \lambda_1, 0; 0, \lambda_2 ] $ but it is more convenient to display them in the vector form as in Equation \ref{MathematicaSDOFeigenvalues}, which is Equation \ref{FirstOrder2DOFeigenvalues}, changed to reflect the subsequent eigenvector term.

\begin{equation} \label{MathematicaSDOFeigenvalues}
\left(%
\begin{array}{c}
  \lambda_1 \\
  \lambda_2 \\
\end{array}%
\right) = \frac{1}{2}\left(%
\begin{array}{c}
  1 + \epsilon + k - \sqrt{\kappa} - 2 \mu^2 - 2 \psi^2 \\
  1 + \epsilon + k + \sqrt{\kappa} - 2 \mu^2 - 2 \psi^2 \\
\end{array}%
\right)
\end{equation}

$ \qquad \quad \kappa \equiv 5 - 2 \epsilon + \epsilon^2 + 2 k - 2 \epsilon k - k^2 $

The 2 DOF eigenvectors in Equation \ref{MathematicaSDOFeigenvectors} are one
dimensional mode shapes for displacement of the hull, $u$, and the armor, $v$, for modes 1 and 2:

\begin{equation} \label{MathematicaSDOFeigenvectors}
\Sigma = \left(%
\begin{array}{cc}
  u_1 & v_1 \\
  u_2 & v_2 \\
\end{array}%
\right) = \frac{1}{2}\left(%
\begin{array}{cc}
  -1 - \epsilon + k + \sqrt{\kappa} & \quad 1 \\
  -1 - \epsilon + k - \sqrt{\kappa} & \quad 1 \\
\end{array}%
\right)^T
\end{equation}

\subsection{Analysis of the 2 DOF SEVP DE}

\label{analysisOf2DOFassumptions}

Mathematica\textsuperscript{\textsc{tm}} shows eigenvalues $ \lambda_i = \omega_i^2 $ for each column $ \Gamma $ of the eigenvector matrix ($ \Sigma $).  Equation \ref{eigenVals1} represents a fixity extreme, a low frequency ``DC" limit $\psi = 0$.



\begin{equation} \label{eigenVals1}
\overrightarrow{\lambda}_{\psi=0} = \frac{1}{2} \left(%
\begin{array}{c}
  1 + \epsilon + k -2\mu^2 - \sqrt{\kappa} \\
  1 + \epsilon + k -2\mu^2 + \sqrt{\kappa} \\
\end{array}%
\right)
\end{equation}

The radical $\kappa$ in this first order correction to the linear eigenvalues (defined for Equation \ref{MathematicaSDOFeigenvalues}) is only a function of stiffnesses.  Compared to prior Mathematica results, this $\psi=0$ model provides a ``DC"  eigenvalue solution that is otherwise not available.

\begin{equation} \label{eigenVectors1}
2\Sigma^T =
\left(%
\begin{array}{cc}
  \overrightarrow{\Gamma_1} & \overrightarrow{\Gamma_2} \\
\end{array}%
\right)^T = \left(%
\begin{array}{cc}
  -1 - \epsilon + k + \sqrt{\kappa} & \quad 1 \\
  -1 - \epsilon + k - \sqrt{\kappa} & \quad 1 \\
\end{array}%
\right)
\end{equation}

\textbf{This is only valid for $\mu_1 = \mu_2$ and $\psi_1 = \psi_2 $ as described earlier.} Otherwise more nonlinear terms remain and the system is not susceptible to SEVP solution for modes in Equation \ref{eigenVectors1}.

\subsection{Synthesis of the 2 DOF SEVP DE}

\label{SEVP2DOFdiscussion}

First we can synthesize the `system' matrix to which these eigenvectors belong.  The kernel of the SEVP is $ [A_1] = |[A]-\lambda [I]| $. Using the fact that $\psi^2$ is the dimensionless frequency, the approximate system matrix A is available from the terms of $ [A_1] $, defined for Equation \ref{MathematicaSDOFeigenvalues}. This is only approximate because of the many combinations of nonlinear and approximately linear variables (like this `linear' $\psi$) selected in Dr. Major Winthrop's dissertation \cite{phdth:Winthrop04} and used in the ``Mathematical Preliminaries" part of the author's CSC thesis \cite{MSthesis:kobold06}.  Taking the matrix $ [A_1] $ we can add a diagonal of $ \lambda_i [I] = \psi_i^2 [I] $ to extract \textit{A} from $ [A_1] = |[A]-\lambda [I]| $.

\begin{equation} \label{synthesize2DOF1}
[A] = \left(%
\begin{array}{cc}
  1 -\mu^2 + \epsilon & -1 \\
  -1 & 1 -\mu^2 + k_{\mbox{open}} \\
\end{array}%
\right)
\end{equation}

Equation \ref{synthesize2DOF1} is essentially a 2 DOF system with springs of stiffnesses k and $\epsilon$, and uniform damping of the same magnitude for both DOF's.  Eigenvectors and eigenvalues are easily recognized in relation to this standard EVP in Equation \ref{synthesize2DOF2} using the assumption ($ \lambda_i = \psi^2 \quad \forall i \in [1, 2] $).

\begin{equation} \label{synthesize2DOF2}
\left(%
\begin{array}{cc}
  1 + -\mu^2 + \epsilon & -1 \\
  -1 & 1 -\mu^2 + k_{\mbox{open}} \\
\end{array}%
\right) \times \overrightarrow{\Gamma_i} %
\\
 = \lambda_1 \times
\overrightarrow{\Gamma_i}
\end{equation}


\subsection{Synthesis of unmatched DE$(\psi_k, \mu_i) \forall i\ne k$}



For clarification and to bring us back to the full nonlinear D.E., use the proviso that the eigenvalues for each DOF are the same for each mode.  Except here we use the physical argument leading to Equation \ref{nonlinearSolution} to maintain thefundamental frequencies equal.  Therefore, the frequency subscript represents the same subscript as that for the eigenvector, rather than matching the damping modes as done in the prior subsection.  But first, define \textit{u}.

The control law, $u(x_1)$ of Equation \ref{controLawFor2DOF} is the nonlinear stiffness \cite[Fig 11]{MSthesis:kobold06}, based on the relative displacement, $ \xi_1 = L^\ast x_1 $. The arctangent switch occurs between the hull lumped point mass and the HRA plate lumped
point mass.

\begin{equation} \label{controLawFor2DOF}
u(\xi_1) = \frac{1}{2}- \arctan(k_x \xi_1)
\end{equation}

The phase  $ \phi_k = \psi_k t  +  \beta(t) \approx \psi_k t $  relates to the eigenvector whose temporal dynamics $ \lambda_i $ describes both DOF's, the hull and HRA point masses. (Here the frequencies $ \psi_k $ are not matched to the damping $ \mu_i $.)  Redefining $\Pi_i$ and $\Omega_i$ for this mixed-index format:

$ \forall \quad \Pi_{i,k} \equiv (\mu_i^2 - \psi_k^2) + 2\mu_i\psi_k\arctan \phi_k,$

$ \qquad \Omega_{i,k} \equiv 2\mu_i \bigg[-\mu_i - \psi_k\arctan \phi_k \bigg] $

\begin{equation} \label{synthesize2DOFphysical1}
\begin{array}{l}
 \qquad \quad \left(%
\begin{array}{cc}
  \Pi_{1,k} & 0 \\
  0 & \Pi_{2,k} \\
\end{array}%
\right) \times \left(%
\begin{array}{c}
  \Lambda_1 \\
  \Lambda_2 \\
\end{array}%
\right)_k + \\
 \qquad \quad \left(%
\begin{array}{cc}
  \Omega_{1,k} & 0 \\
  0 & \Omega_{2,k} \\
\end{array}%
\right) \times \left(%
\begin{array}{c}
  \Lambda_1 \\
  \Lambda_2 \\
\end{array}%
\right)_k  + \\
  \left(%
\begin{array}{cc}
  1 + \epsilon u(x_1) & -1 \\
  -1 & k_{\mbox{open}} \\
\end{array}%
\right) \times \left(%
\begin{array}{c}
  \Lambda_1 \\
  \Lambda_2 \\
\end{array}%
\right)_k  = \left(%
\begin{array}{c}
  0 \\
  0 \\
\end{array}%
\right) \\
\end{array}
\end{equation}

Equation \ref{synthesize2DOFphysical1} provides an iid estimate of the dynamics modelled as N independent identically distributed oscillators.  Since `oscillators' for each DOF in a FE model and in continuous media (reality) are not independent nor of a random distribution in a structure, this relationship merely guides independent behavior that quickly effects other DOF's.  More importantly, the ``nonlinear" eigenvalue behavior in Figure \ref{realEigen2dof} shows how the symmetrical modes are affected by stiffness change \cite[p. 148]{MSthesis:kobold06}.

A Lyapunov function form of energy balance is defined in the author's CSC thesis \cite[App F]{MSthesis:kobold06} that serves to help validate the stability of SDOF and 2DOF relations.

\bibliographystyle{ieeetran}


\bibliography{modLobsrv}


{}
%

\end{document}